\documentclass[12pt,preprint]{aastex6}
\usepackage{exscale}
\usepackage{Picins}
\usepackage{color}
\usepackage{rotating}
\usepackage{graphicx}
\usepackage{epstopdf}
\usepackage{float}
\usepackage{amsfonts}
\usepackage{amssymb}
\usepackage{hyperref}
\usepackage{natbib}
\usepackage{caption}
\usepackage{subfigure}
\usepackage{overpic}
\usepackage{ulem}
\usepackage{xcolor}
\usepackage{multirow}
\usepackage{array}
\usepackage{booktabs}
\usepackage{appendix}

\shorttitle{Extensive Study of a Coronal Mass Ejection with UV and WL coronagraphs}
\shortauthors{Ying et al.}

\begin{document}
\title{Extensive Study of a Coronal Mass Ejection with UV and WL coronagraphs: the need for multi-wavelength observations}
\author{Beili Ying\altaffilmark{1,2}, Alessandro Bemporad\altaffilmark{3}, Li Feng\altaffilmark{1}, Lei Lu\altaffilmark{1}, Weiqun Gan\altaffilmark{1}, Hui Li\altaffilmark{1}}

\altaffiltext{1}{Key Laboratory of Dark Matter and Space Astronomy, Purple Mountain Observatory, Chinese Academy of Sciences, 210023 Nanjing, China}
\altaffiltext{2}{School of Astronomy and Space Science, University of Science and Technology of China, Hefei, Anhui 230026, People's Republic of China}
\altaffiltext{3}{INAF-Turin Astrophysical Observatory, via Osservatorio 20, I-10025 Pino Torinese (TO), Italy}
\email{yingbl@pmo.ac.cn;lfeng@pmo.ac.cn}
\begin{abstract}
Coronal Mass Ejections (CMEs) often show different features in different band-passes.  By combining data in white-light (WL) and ultraviolet (UV) bands, we have applied different techniques to derive plasma temperatures, electron density, internal radial speed, etc, within a fast CME. They serve as extensive tests of the diagnostic capabilities, developed for the observations provided by future multi-channel coronagraphs (such as \textit{Solar Orbiter}/Metis, \textit{ASO-S}/LST, \textit{PROBA-3}/ASPIICS).
The involved data include WL images acquired by SOHO/LASCO coronagraphs, and intensities measured by SOHO/UVCS at 2.45 R$_{\odot}$ in the UV (H {\small I} Ly$\alpha$ and O {\small{VI}} 1032 \AA\ lines) and WL channels. Data from the UVCS WL channel have been employed for the first time to measure the CME position angle with polarization-ratio technique. Plasma electron and effective temperatures of the CME core and void are estimated by combining UV and WL data. Due to the CME expansion and the possible existence of prominence segments, the transit of the CME core results in decreases of the electron temperature down to $10^{5}$ K. The front is observed as a significant dimming in the Ly$\alpha$ intensity, associated with a line broadening due to plasma heating and flows along the line-of-sight. The 2D distribution of plasma speeds within the CME body is reconstructed from LASCO images and employed to constrain the Doppler dimming of Ly$\alpha$ line, and simulate future CME observations by Metis and LST.
% \sout{The electron temperature ratio of the CME manifested the obvious heating in the CME void and the cooling in the core.}
\end{abstract}
\keywords{Sun: corona $-$ Sun: coronal mass ejections (CMEs) $-$ Sun: UV radiation}

\bibliographystyle{apj}
\section{Introduction}
Coronal Mass Ejections (CMEs), often associated with solar flares and/or filament eruptions \citep{Webb2012,Morgan2012}, are one of the most dramatic phenomena occurring in the solar atmosphere. CMEs can release a large amount of energy ($10^{29}-10^{32}$ erg) and magnetized plasma ($10^{15}-10^{16}$ g) with fast speeds even up to 3500 $\rm km~s^{-1}$ \citep{Gopalswamy2009}. In the low corona, many studies believe that a CME may originate as a magnetic flux rope, which contains a magnetic structure with a coherent magnetic field winding around its central axis \citep{Chen1989,Zhangjie2012}, and with high temperature plasmas \citep{Reeves2011,Cheng2014a,Ying2018}. As CMEs propagate into the interplanetary space, they often take the form of magnetic clouds, varying structures of twisted magnetic flux accompanied by low pressure plasma \citep{Schwenn2006}. When the speed of a CME is larger than the local fast magnetosonic speed, a shock can be generated, often accompanied with a type-II radio burst \citep{Gopalswamy2005,Gopalswamy2006,Gopalswamy2010}. Many of the Solar Energetic Particles (SEPs) associated with CMEs are believed to be produced as the shock passes through the corona \citep{Schwenn2006}, but possible accelerations occurring in solar flares are still debated \citep[see review by][and references therein]{Reames2013}. When CMEs encounter the Earth's magnetosphere, the direct bombardment of SEPs and the disturbances of the geomagnetic field can disrupt power grid and disable satellites \citep{Lanzerotti2001}.

In order to understand physical mechanisms responsible for solar activities and to monitor solar eruptions, including CMEs and flares, many ground-based and space-based instruments in different wavelengths have been used to observe the different layers of the solar atmosphere. Daily EUV images of the solar disk were and are being provided by the Extreme ultraviolet Imaging Telescope (EIT) \citep{Delab1995} onboard the \textit{Solar and Heliospheric Observatory (SOHO)}, the Transition Region and Coronal Explorer \citep[TRACE;][]{Handy1999}, and more recently by the Atmospheric Imaging Assembly \citep{Lemen2012} onboard the \textit{Solar Dynamics Observatory (SDO)}. In the white-light (WL) band, images observed by many coronagraphs can be utilized to study physical parameters of CMEs in the corona, including the Large Angle Spectroscopic Coronagraph \citep[LASCO;][]{Brueckner1995} onboard the SOHO mission, and the COR1 and COR2 coronagraphs \citep{Howard2008} onboard the \textit{Solar TErrestrial RElations Observatory (STEREO)} mission in the space. The UV Coronagraph Spectrometer \citep[UVCS;][]{Kohl1995} on SOHO, with a field of view (FOV) of 42' (projected slit length), also permitted to investigate great details of CMEs by using different UV emission lines. Main lines include the H {\small I} Ly$\alpha$ 1216 {\AA}, and O {\small VI} $\lambda\lambda$1032/1037 {\AA} doublet spectral lines.

Over past years, many investigators combined the WL coronagraphic images and the UV observations from UVCS to determine different parameters of CMEs, including the distribution of plasma temperatures (electron and ion), the evolution of the CME core and front, and elemental distributions \citep{Akmal2001, Raymond2003, Bemporad2007, Ciaravella2003, Ciaravella2006}. With UVCS many CME-related phenomena were also investigated, such as post-CME current sheets \citep{Bemporad2008,Cai2016}, CME-driven reconnections \citep{Bemporad2010b}, CME-driven shocks \citep{Ciaravella2005, Ciaravella2006, Bemporad2010}, as well as SEPs' accelerations in CME-driven shocks \citep{Ciaravella2005}. The first stereoscopic and spectroscopic reconstruction of a CME has been performed by the combination of the UVCS spectra and STEREO WL images \citep{Susino2014}. All the plasma physical properties could not have been derived in these works from standard WL coronagraphic images alone. For this reason, in the near future several new instruments have been designed to provide simultaneous observations of solar corona in the WL bands and UV spectral lines. One of these instruments is the Metis coronagraph \citep{Antonucci2017,Fineschi2020} onboard the \textit{ESA-Solar Orbiter} mission (launched on February 10, 2020) with field-of-view (FOV) of $1.6^{\circ}-2.9^{\circ}$ (corresponding to projected altitudes from 1.7 to 3.1 $\rm R_{\odot}$ when the spacecraft reaches the closest approach at 0.28 AU). Similar to Metis, the Lyman-alpha Solar Telescope \citep[LST;][]{Li2019,Feng2019} instrument onboard the future \textit{Chinese Advanced Space-based Solar Observatory (ASO-S)} mission \citep[][to be launched in 2022]{Gan2019} can image the Sun and the inner corona with a FOV up to 2.5  $\rm R_{\odot}$  in both WL and H {\small I} Ly$\alpha$ line at high temporal and spatial resolutions. These new instruments will lead to new insights about the physical processes including CME heating and acceleration, solar wind acceleration, as well as SEP acceleration.

Many advanced scientific preparations have been done for these new instruments. The plasma temperature and density are vital parameters to study the evolution of CMEs. \citet{Susino2016} have combined the WL images from LASCO/C2 and the UV H {\small I} Ly$\alpha$ line intensities from the UVCS spectrometer to demonstrate how CME plasma electron densities and temperatures can be derived even without slit-spectroscopic data. These diagnostic techniques will be applied to the future observations from the new coronagraphs. Furthermore, \citet{Bemporad2018} used a numerical three-dimensional (3D) magnetohydrodynamic (MHD) simulation of a CME, and subsequently obtained the synthetic WL and UV (in Ly$\alpha$ line) images to evaluate the capability of electron temperature estimation from the diagnostic method. According to the formation mechanism of the H {\small I} Ly$\alpha$ line, the Doppler-shift of the scattering profile with respect to the disk profile results in a less efficient atomic excitation, hence leading to the so-called Doppler dimming \citep{Beckers1974, Withbroe1982}. It plays a vital role in this diagnostic method, and is directly affected by the radial velocity distributions of the CMEs. In order to take into account this effect in future CME images in the Ly$\alpha$ line, \citet{Ying2019} showed how the cross-correlation technique can be employed to derive the two-dimensional (2D) radial speed map of a CME, and thus the distribution of H {\small I} Ly$\alpha$ Doppler dimming factors.

In this work, the main purpose is to extensively test the diagnostic tools for CME parameters, e.g., plasma temperatures, electron density, internal radial speed, and to emphasize the importance to combine observations in the WL and UV band-passes. We analyzed a fast CME with a CME-driven shock, passing through the FOV of the UVCS slit and captured by the LASCO coronagraph at the same time. This work is a follow-up of \citet{Susino2016}, but in that work the CMEs propagated with slow speeds (no more than 300 $\rm km~s^{-1}$), and the Ly$\alpha$ intensities were not significantly affected by Doppler dimming. In this work, we show how the same techniques can be applied to a fast CME $\sim 1100~\rm km~s^{-1}$. Moreover, given the low cadence of available images for this event, we provide an alternative simple geometrical technique, which is still applicable in case the cadence of future data is not sufficient to apply cross-correlation analysis shown by \citet{Ying2019}, to derive the 2D radial velocity map of the CME, and further to better constrain the Ly$\alpha$ Doppler dimming effect. The analysis of UVCS data allowed us to measure across the CME the evolution of both electron and effective temperatures. This work also analyzed for the first time the sequence of data acquired during a CME by the UVCS WL Channel (WL channel), and employed these data to measure the CME's position angle with respect to the plane-of-sky (POS) with the polarization-ratio technique.

The paper is organized as follows: we describe the EUV image and LASCO observations in \autoref{sec:CME_WL} and show the UV and WL channel observations from UVCS in \autoref{sec:CME_UVCS}. The plasma diagnosis of the CME are described in \autoref{sec:CME_tmperature}. Finally, we discuss and conclude our results in \autoref{sec:discussion_conclusion}.

%%%%%%%%%%%%%%%%%%%%%%%%%%%%%%%%%%%%%%%%%%%%%%%%%%%

%%%%%%%%%%%%%%%%%%%%%%%%%%%%%%%%%%%%%%%%%%%%%
\section{CME in the LASCO and EUV images}
\label{sec:CME_WL}
\begin{figure}
    \centering
    \includegraphics[width=1.\textwidth]{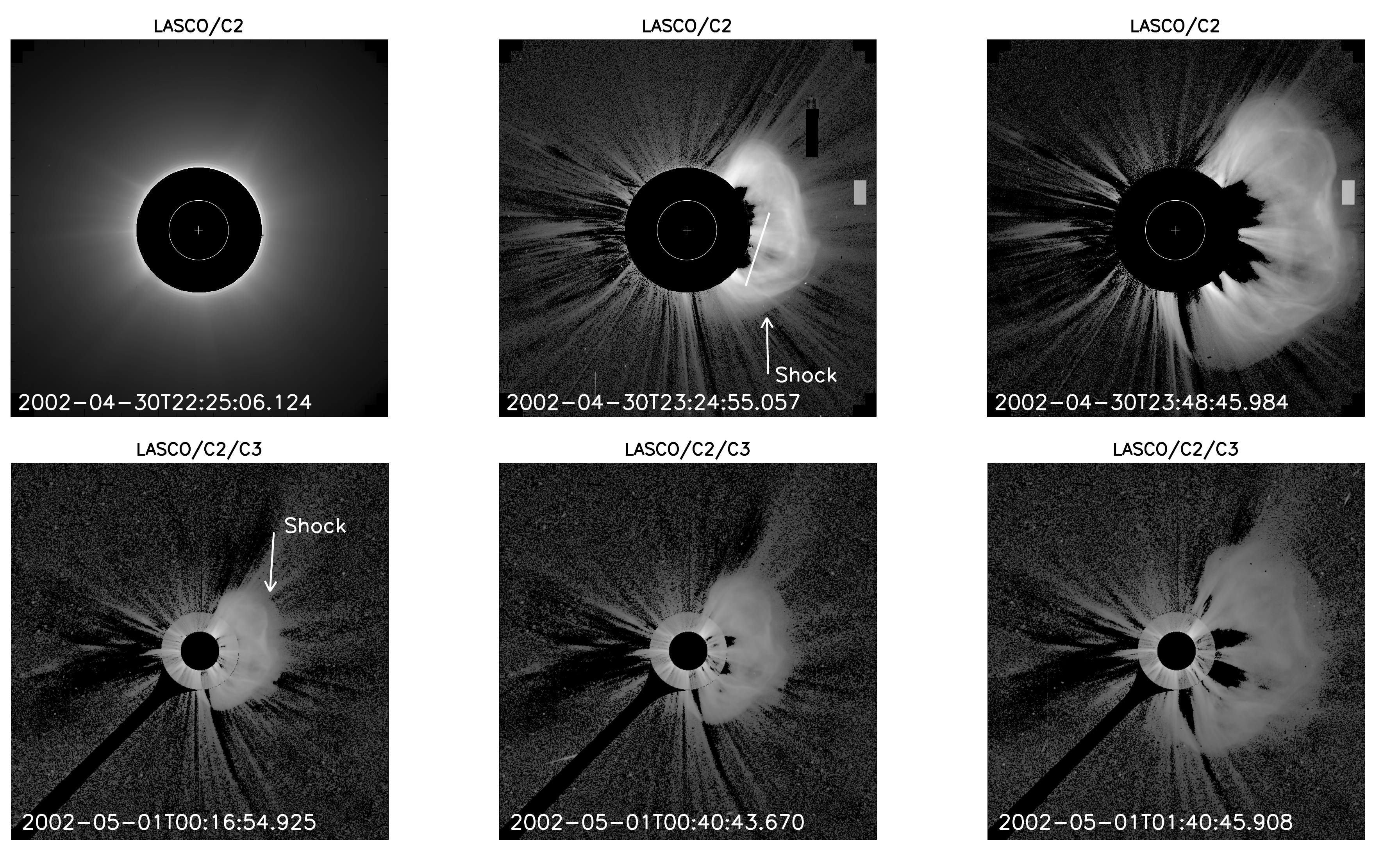}
    \caption{Top: the left panel is the pre-event background from the LASCO/C2 at 22:25 UT. Two LASCO/C2 base-difference images acquired at 23:24 UT and 23:48 UT are shown in the middle and right panels. In the middle panel, the SOHO/UVCS field of view (represented as a white line) is also superimposed to mark the portion of the solar corona sampled by the spectrometer during the transit of the CME . Bottom: evolution of the CME in the FOVs of LASCO/C2 and C3 from 00:17 UT to 01:41 UT on 2002 May 1. Shock signatures are marked by white arrows. White circles and plus signs in the top panels denote the solar limb and its center .}
    \label{fig:c23_img}
\end{figure}

\begin{figure}
    \centering
    \includegraphics[width=1.\textwidth]{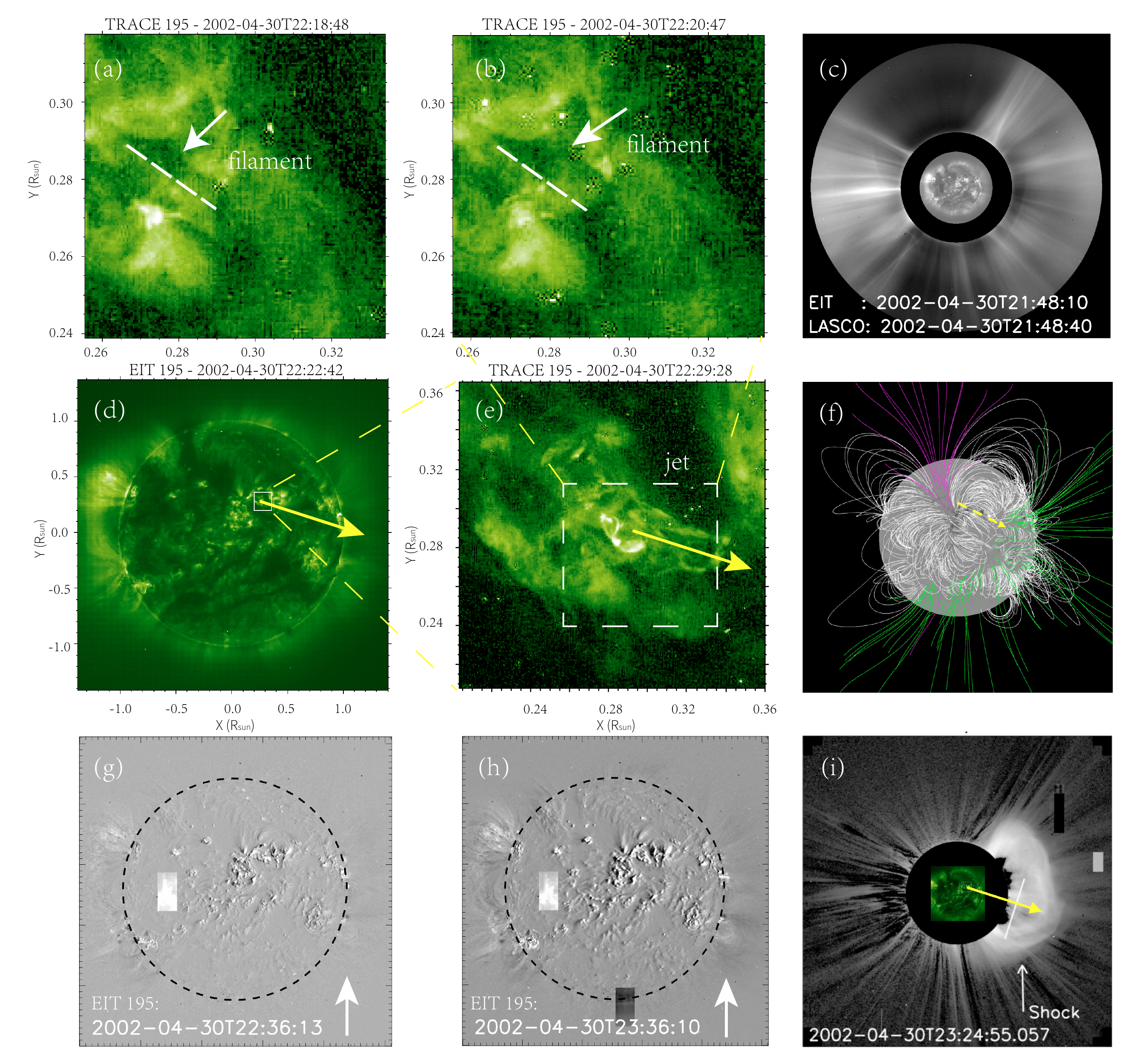}
    \caption{(a)-(b): TRACE 195 {\AA} images at $\sim$22:19 UT and $\sim$22:21 UT. The rising filament-like structure is marked by white arrows and the dashed lines. (c): the EIT 195 {\AA} image combined with the LASCO/C2 image (processed with the NRGF) acquired at 21:48 UT, before the CME eruption. (d): the EIT 195 {\AA} image at 22:23 UT, showing the location of the only eruption on the solar disk that was observed by TRACE (marked by the white box). (e): the TRACE 195 {\AA} image at 22:29 UT, and the location of jet; yellow arrows have the same inclination in panels (d), (e) and (f), indicating that the projected direction of the jet is consistent with the propagation direction of the CME nose as indicated in panel (i). The area covered by panel (e) is defined by the white box in panel (d), while panels (a) and (b) show the area zoomed in panel (d). (f): coronal magnetic field lines extrapolated by PFSS method over-plotted in MDI magnetogram. Purple and green lines denote the open magnetic field lines with negative and positive magnetic fields. A dashed yellow arrow shows the projected inclination of the loop system overlying the jet source region. (g)-(h): two EIT 195 {\AA} base-difference images at 22:36 UT and 23:36 UT, obtained after subtracting the pre-event image at 21:48 UT. White arrows indicate the CME dimming region and post-CME loops. (i): the LASCO/C2 base-difference image combining the EIT 195 {\AA} image acquired at $\sim$23:25 UT.}%Top: SOHO/EIT 195 {\AA} images acquired before and during the CME eruption on 2002 April 30. The observed times of these images are from 21:48 UT to 23:36 UT. Bottom: left panel shows the EIT image combining with the LASCO/C2 image (processed with the NRGF) at 21:48 UT. The display of the last two base-difference images subtracting the pre-event background at 21:48 UT is to show the little change of the active regions on the disk, which might imply that the source region of the eruption is behind the solar disk. Dashed line represents the solar limb.}
    \label{fig:eit_img}
\end{figure}

In the event of 2002 April 30, a fast, partial-halo,  CME appeared in LASCO/C2 images showing a clear front, but without a clear core. The CME was first captured by the LASCO/C2 at $\sim$23:25 UT with an estimated POS speed of $\sim 1100 ~\rm km~s^{-1}$ at the CME nose (\autoref{fig:c23_img}). Before the appearance of the CME, there exist some faint streamer-like structures in the LASCO/C2 images as shown in \autoref{fig:eit_img} (c). As the CME propagated into the FOV of the LASCO/C3, also no clear signature of a bright core was detected in the WL images.

The identification of the source region for this event is not certain. On 2002 April 30, SOHO/EIT acquired full-disk images in the 195 {\AA} filter with a time cadence by 12 minutes, but no clear flare, EUV dimming, or post-CME loops were observed by EIT in the visible hemisphere as shown in \autoref{fig:eit_img} (g) and (h). Higher-cadence EUV images were acquired with the same 195 {\AA} filter by TRACE, and the instrument was pointing on a North-West active region group, with a FOV limited to an area by 0.4 R$_\odot \times$ 0.4 R$_\odot$. \autoref{fig:eit_img} (e) is a zoomed-in TRACE 195 {\AA}  image. Quite interestingly, this region shows the occurrence of a small scale jet-like eruption (in the white box of \autoref{fig:eit_img} (e)) starting from 22:25 UT, which is just 9 minutes before the CME starting time as provided by the automatic LASCO CDAW catalog. The source region of this small eruption is located at N15$^{\circ}$ and W16$^{\circ}$, as shown in \autoref{fig:eit_img}.  Before the eruption, a filament-like structure (marked by white arrows) appearing as a dark moving plasma feature (hence seen in absorption), starts to rise from 22:10 UT, as shown in \autoref{fig:eit_img} (a) and (b). We find that the propagation of this jet is highly twisted. It is non-radially propagated with a clear westward inclination (\autoref{fig:eit_img} (d)). Comparing panels (d) and (i) in \autoref{fig:eit_img}, it is clear that the projected propagation direction of the jet is in agreement with the CME nose direction.

Due to the good temporal and directional correlations between the jet and the CME, this could be the CME source region. For the further confirmation, we investigated the distribution of coronal magnetic field lines as provided by Potential Field Source Surface (PFSS) extrapolations \citep{Schrijver2003}, and found a system of closed field lines, with foot-points rooted near the small jet source and with a South-West inclination matching the direction of the jet eruption (dashed yellow arrow). If the twist injected by the jet into the loop system is large enough, the loops may be destabilized and ascend to become a CME. However, in this interpretation the CME width should be restricted by the loop system, which may not be as wide as the partial-halo CME that we see in LASCO. Many studies have revealed that jets can trigger CMEs \citep{Liujiajia2015,Panesar2016,Ying2018}, but these jet-triggered CMEs are usually narrow, and they are not halo or partial-halo CMEs. What's more, we cannot rule out the possibility that the correspondence between times and directionalities of the jet and the CME are just a coincidence.

Alternatively, the CME real source region could be also located behind the visible solar hemisphere. The EIT 195 {\AA} images show little evidence of an eruption which happened above the west limb of the solar disk from southern hemisphere starting from 21:48 UT to 23:36 UT. Two base-difference images are shown in \autoref{fig:eit_img} (g) and (h), with pre-event background subtraction at 21:48 UT. Especially, \autoref{fig:eit_img} (h) clearly shows a post-CME dimming and post-CME loops (marked by a white arrow) that are both located off-limb without any other signature on-disk, and many active regions corotating with solar rotation crossed the disk in the days before the CME. This means that the CME source region was also possibly located just behind the visible hemisphere. Due to the insufficient evidences and observations, it is not possible to identify the most probable source of this event.

The speed of the CME suggests that a fast-mode shock could be formed, hence a shock signature might be observed in WL images. \citet{Vourlidas2003} first revealed that a compressed shock region can be strong enough to be detected by the LASCO coronagraph, and appear as a faint front preceding the CME bright leading edge. In this event, the faint front region was observed and denoted by white arrows in the LASCO images at 23:25 UT of 2002 April 30 and 00:17 UT of the next day in \autoref{fig:c23_img}. On the other hand, no clear signature of a type-II radio burst was observed by Wind/WAVES instrument. 

%Type II solar radio bursts are generally considered to be radio signatures of shock waves, with slowly drifting structures from higher frequencies to lower frequencies identified in dynamic spectra. Langmuir waves are yielded by accelerated electrons at the shock front, subsequently converting to electromagnetic waves near the electron plasma frequency $f_p$ and 2$f_p$.
%Complex multi-lane structures of type II radio bursts were captured in the radio dynamic spectra staring from $\sim$ 22:10 UT in \autoref{fig:c23_img} (top right). However, the signatures is faint and weak at the beginning (marked by the left arrow), which might be due to the low signal-to-noise ratio and the non-linear formation process of the shock.%
%%%%%%%%%%%%%%%%%%%%%%%%%%%%%%%%%%%%%%%%%%%%%%%%
\section{CME in the UVCS observations} 
\label{sec:CME_UVCS}
This event was observed by the UVCS slit both in the UV range (with the so-called O {\small VI} channel) and the WL range. The CME crossed the FOV of the UVCS instrument acquiring data with the slit center located at polar angle (PA, measured counter-clockwise from north polar) of $252^{\circ}$ (corresponding to 18$^\circ$ SW), and projected altitude of 2.45 $\rm R_{\odot}$. This position is marked by a white solid line in \autoref{fig:c23_img} (top middle panel), and it shows the projected FOV of the UV channel, while the FOV of the WL channel consists of a single pixel located at the slit center. The UVCS observations covered the time interval from 19:43 UT on 2002 April 30 to 01:55 UT on May 1st, and this time interval included both the pre-event corona and the CME eruption. The projected width of the slit is $21^{"}$ and the spatial bin size is $42^{"}$ in this case (6 pixels per bin). All data were calibrated and reduced by the latest available version of the UVCS Data Analysis Software (DAS 51).%%

\subsection{UV channel}

For these observations, the detected spectral range in the UV channel includes the H {\small I} Ly$\alpha$ $\lambda$1215.6 {\AA} line obtained from the redundant path of the O {\small VI} channel with spectral binning of 1 pixel (0.0915 {\AA}), and the O {\small VI} $\lambda\lambda$ 1031.9/1037.6 {\AA} doublet with spectral binning of 1 pixel (0.0991 {\AA}). The exposure time for each spectrum is 10 minutes.

\begin{figure}
    \centering
    \includegraphics[width=1.03\textwidth]{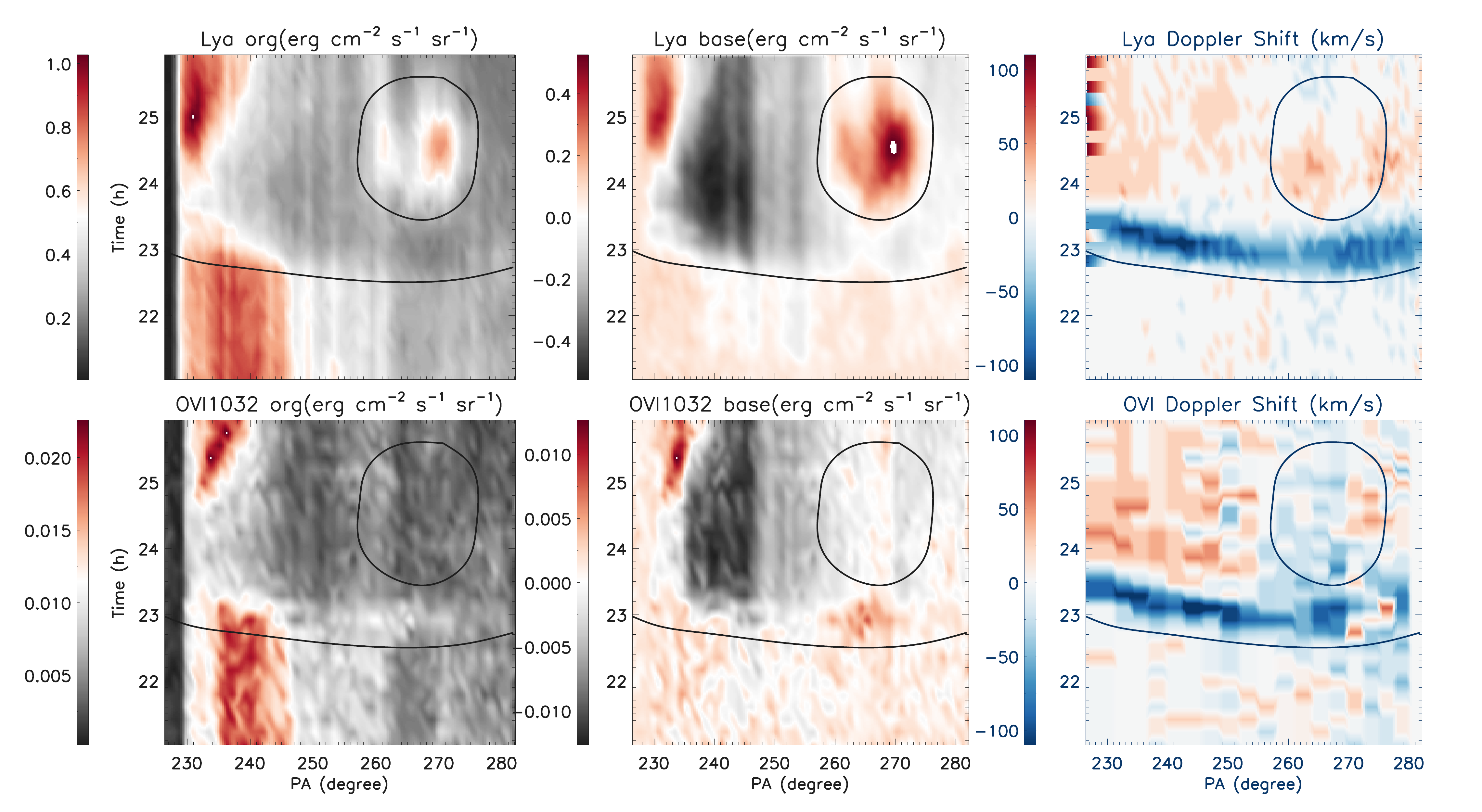}
    \caption{Top: the evolution of H {\small I} Ly$\alpha$ total intensity (left), base-difference image (middle), and Doppler shift speed (right). Bottom: the same evolution as that of the top panels but in O {\small VI} $\lambda$1032 {\AA} line. Time in the y-axes is given starting time at 21:00 UT on 2002 April 30. The x-axes show the polar angle of the UVCS spectrometer slit. A core and a front of the CME are marked by black lines.}
    %the evolution of O {\small VI} $\lambda$1032 {\AA} total (left) intensity, base-difference image (middle), as well as its Doppler shift speed (right)
    \label{fig:uvcs_img}
\end{figure}

\begin{figure}
    \centering
    \includegraphics[width=0.4 \textwidth]{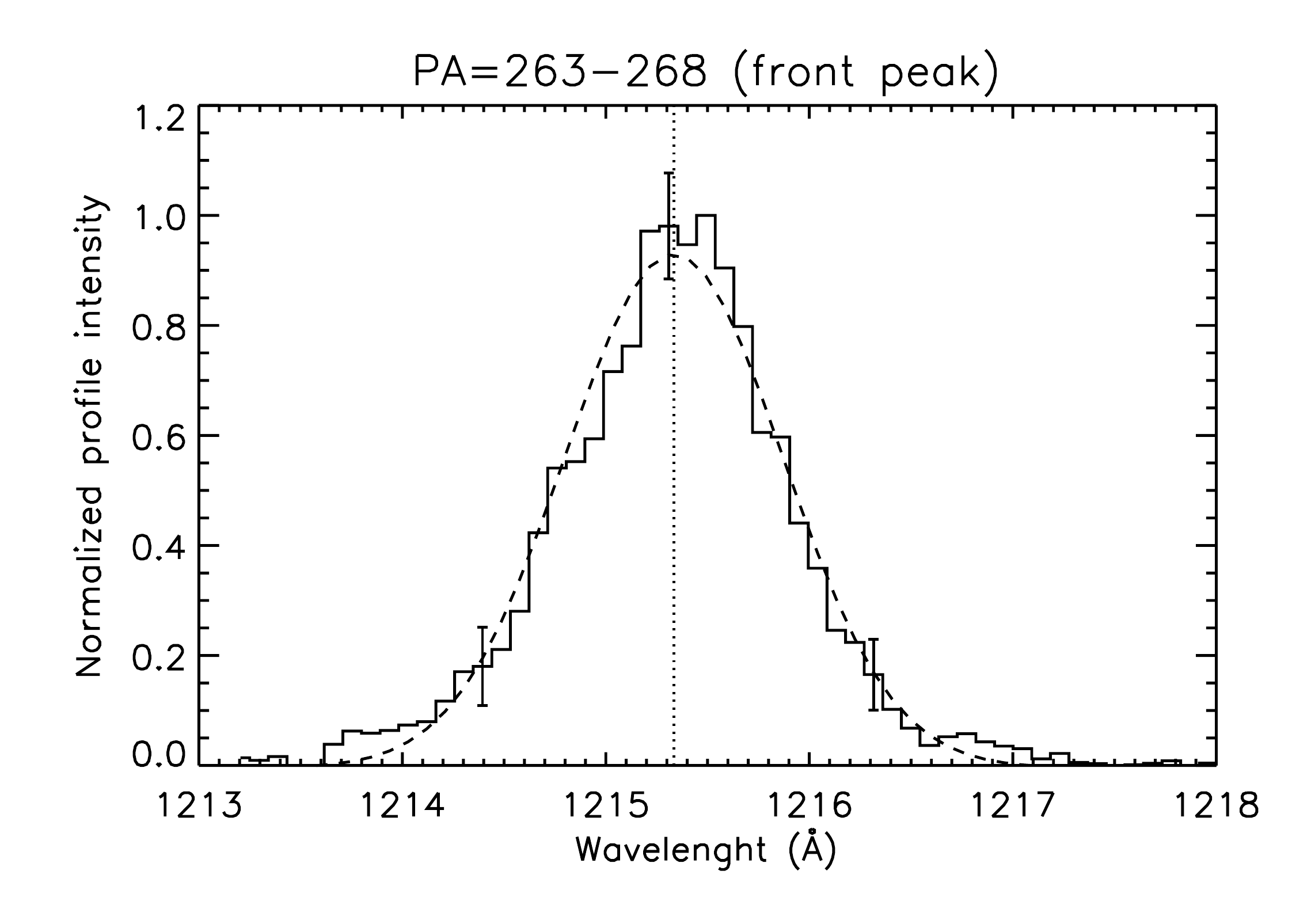}
    \caption{A Ly$\alpha$ normalized line profile (solid line) averaged over $5^{\circ}$ and the corresponding single-Gaussian fit (dashed line) at 22:55 UT and $\rm PA=263^{\circ}-268^{\circ}$ ( marked by a arrow in the left panel of \autoref{fig:shock}). A vertical dotted line shows the centroid of the profile. Error bars of the normalized profile are propagated from the standard deviations of the average of the Ly$\alpha$ line profiles over $5^{\circ}$.}
  \label{fig:profile}
\end{figure}

\autoref{fig:uvcs_img} shows the time evolution ($y-$axis) of the UV intensities measured along the UVCS slit ($x-$axis) in the H {\small{I}} Ly$\alpha$ (top left panel) and O {\small{VI}} $\lambda$1032 {\AA} (bottom left panel) lines, as well as corresponding Doppler shift speeds (right panels) derived from Gaussian fitting of these two lines. In \autoref{fig:uvcs_img}, the horizontal axis corresponds to PA along the UVCS FOV, and the vertical axis is the observation time starting from 21:00 UT on April 30. The left columns of \autoref{fig:uvcs_img} show the evolution of the total UV intensities without pre-event intensity subtraction, while the middle columns show ``base-difference" images of the Ly$\alpha$ and O {\small{VI}} $\lambda$1032 {\AA} intensities after subtractions of the average pre-event latitudinal intensity distributions, to better show the fainter CME structures. In order to display the overall distributions of the Doppler shift speeds, we apply the single-Gaussian fit to simplify the analysis process, because majority of profiles of the H {\small{I}} Ly$\alpha$ line are almost symmetric. \autoref{fig:profile} shows an example of the Ly$\alpha$ normalized line profile averaged over $5^{\circ}$ at 22:55 UT and $\rm PA=263^{\circ}-268^{\circ}$. The Doppler shifts during the CME have been measured with respect to the centroid of the pre-CME coronal profiles, to avoid any possible error related with UVCS instrumental wavelength calibration. The Doppler shift images display a strong blue-shift with a speed of $\sim 100 \rm~km~s^{-1}$ in the CME front, while the red-shift in the core is very weak in Ly$\alpha$ and almost absent in the O {\small{VI}} line, which means that the CME is mostly blue-shifted and moves with a velocity component towards the observer. The red-shift of the CME core in Ly$\alpha$ line might be due to the asymmetry of CME expansion. As we mention above, the source region of the CME was possibly located on or behind the visible hemisphere. It is possible that a back-side CME owns a blue-shift speed in the front, as shown for instance by \citet{Ciaravella2006} who reported a back-side CME with a blue-shifted emission in the front corresponding to a speed of 240 $\rm km~s^{-1}$. Through a statistical research on halo CMEs, \citet{Ciaravella2006} have found that if the centroid Doppler speeds of halo CME fronts are smaller enough than the POS speed,  the CME fronts would be swept-up coronal plasma rather than material carried in magnetic loops, supporting geometries of the CME fronts being ``ice cream'' cones.
%\textbf{In fact, the line profiles of the CME front and core could be asymmetry. The usage of the double-Gaussian fit can to separate emissions from the CME and coronal background, and further allow us to infer more information of the CME, such as the CME shape, propagation, and temperature \citep{Ciaravella2006}. Thus, we also explore the Ly$\alpha$ line profile shapes of the CME front and core at the fixed PA range, and the CME Ly$\alpha$ normalized line profiles with double-Gaussian fits are displayed below.}

In the FOV of UVCS, the Ly$\alpha$ base-difference image (top middle) shows a bright core part (surrounded by the black line), and a dimming front (shown by the black curved line), while in the O {\small{VI}} intensity (the bottom middle panel of \autoref{fig:uvcs_img}) the front appears as a faint intensity increase. On the other hand, the WL images in the LASCO coronagraph display that the CME owns a bright leading edge, but no clear core structure all the time. Therefore, this event shows a very different appearance in different bands: a bright core and a dimmed front in the UV Ly$\alpha$ $\lambda$1215.6 {\AA} line, a slightly bright front and a faint core in the UV O {\small{VI}} $\lambda$1032 {\AA} line, a bright front and not clear core in the WL images. Comparison between the CME appearance in the UV and WL will be further discussed in \autoref{sec:com_uv_wl}. Here we remind that, these differences are related to the different physical mechanisms involved in the emissions in different bands from different CME parts. In addition, the distinct behaviors of these UV line intensities and Doppler shifts might be due to the capture of different parts of the CME plasma, considering that ionization temperatures of O$^{5+}$ ions responsible for O {\small{VI}} emission are higher than those of neutral H atoms emitting Ly$\alpha$. As pointed out by \citet{Kohl2006}, the brightness of a CME core can increase by three orders of magnitude in the Ly$\alpha$ lines over the process of a few minutes when its material reaches the UVCS slit, while the WL intensity increase never exceed a few percent. Apparently, something similar happened for this event where the core corresponds to approximately a 70\% increase in the UV Ly$\alpha$ intensity, and a negligible increase in the WL images. The same behavior was also observed for a rather small CME during the 2017 total solar eclipse with two optical lines of Fe {\small{XI}} and Fe {\small{XIV}} \citep{Boe2020}. This might indicate that the intensities of emission lines in general change much more due to CMEs than do white light. Thus CMEs could motivate more than only UV (such as H {\small I} Ly$\alpha$ and O {\small VI} doublet lines) observations in the corona.
What's more, \citet{Cremades2004} pointed out that in the WL coronagraph observations the CME core cannot be observed if the symmetry axis of CME is perpendicular to the LOS. Thus, lack of a CME core in WL images might also be due to its orientation of symmetry axis.
% \textbf{In this work, the UV Ly$\alpha$ intensity of the CME core is contributed by the collisional excitation ($\propto n_e^2$, electron density) and the radiative excitation ($\propto n_e$), as shown in \autoref{sec:Formation}; meanwhile, the hydrogen element abundance is sensitive to the change of the electron temperature, while the WL brightness only depends on the electron density ($\propto n_e$) due to the Thomson scattering.}

The detection of CME front as a dimming in the UV Ly$\alpha$ line is not a surprise and was already reported before \citep[e.g.,][]{Ciaravella2006,Bemporad2010}. More recently, \citet{Bemporad2018} confirmed with MHD simulation that the CME front in the UV Ly$\alpha$ appears with a relative reduction of coronal emission, while it is bright in the WL image. This difference between the UV and WL is mainly due to the Doppler dimming effect \citep{Bemporad2018} and significant plasma heating going on in the simulated CME front. For this event, the intensity dimming observed after the transit of the front appears with a delay in the O {\small VI} $\lambda$1032 line with respect to the Ly$\alpha$ line. This observation is approximately the same as that reported by \citet{Bemporad2010}. The delay of the intensity reduction of the O {\small VI} $\lambda$1032 line observed after the Ly$\alpha$ dimming is likely due to the enhancement of the collisional component of the O {\small VI} line, because of plasma compression occurring in the CME front.%%

\begin{figure}
    \centering
    \includegraphics[width=1.\textwidth]{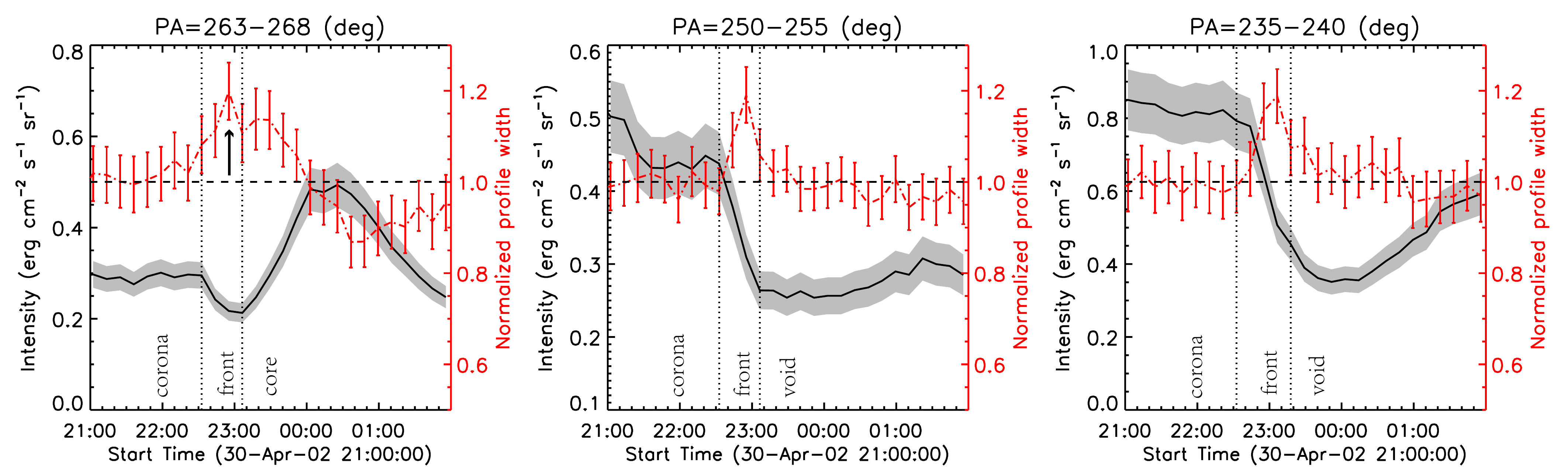}
    \caption{Ly$\alpha$ total intensity (solid line) and normalized profile width (dotted-dashed line) evolution averaged over $\rm 5^{\circ}$ at different PA (please refer to \autoref{fig:uvcs_img} for the location of these angular intervals). The H {\small I} Ly$\alpha$ profile widths are normalized by the average Ly$\alpha$ profile width (marked by the horizontal dash line) of the coronal background. The transits of different parts of the CME are divided by vertical dotted lines. Errors of the normalized profile widths are propagated from the $1\sigma$ uncertainty of the single-Gaussian fitting, and errors of the Ly$\alpha$ total intensity are propagated from the 22\% uncertainty of the UVCS radiometric calibration. A arrow in the left panel indicates the corresponding position and time of the Ly$\alpha$ normalized profile in \autoref{fig:profile}.}
    \label{fig:shock}
\end{figure}

In order to explore the evolution of different stages, we obtain the H {\small I} Ly$\alpha$ profile width normalized by the average profile width of the coronal background at different PA ranges as well. Errors are propagated from the single-Gaussian fit with $1\sigma$ uncertainty. \autoref{fig:shock} displays a comparison between the evolution of the Ly$\alpha$ total intensity (solid line) and the normalized profile width (dotted-dashed line) averaged over $5^{\circ}$ at three different PAs (as indicated at the top of each panel). Errors of the Ly$\alpha$ total intensities are propagated from the UVCS  UV radiometric calibration. \citet{Gardner2002} reported that the UVCS radiometric calibration error in UV is about 20\%-22\% for the first order lines, thus, we assume the UV error is 22\% in this work. Horizontal dashed lines, whose value is equal to 1, denote the normalized profile width of the coronal background. 

According to the evolution of the Ly$\alpha$ total intensity with time and the change of the normalized profile width at PA$=263^{\circ}-268^{\circ}$ (\autoref{fig:shock}, left panel), the transit of this event can be divided into (vertical dotted lines) three phases: the pre-event corona, the CME dimming front, and the CME dense core. The arrival of the CME front corresponds to a Ly$\alpha$ intensity Doppler dimming that could be due to both radial speeds and higher temperatures of the front plasma, accompanied by a Ly$\alpha$ profile broadening mainly due to bulk flow motions occurring along the LOS and related with the CME front expansion, as shown in the left panel of \autoref{fig:shock}.  Later on, the arrival of the CME core results in a Ly$\alpha$ intensity increase (related with higher plasma densities and lower temperatures, hence a higher fraction of neutral H) and a Ly$\alpha$ profile narrowing (due to lower plasma temperatures of plasma embedded in the core). The interpretations above will be further supported by data analysis in the following sections. On the other hand, the middle and right panels of \autoref{fig:shock} (providing, respectively, the evolution at PA$=252^{\circ}$ and $237^{\circ}$ averaged over $5^{\circ}$) show that at these latitudes only the Ly$\alpha$ intensity dimming associated with the CME front are captured by the UVCS slit. In any case, all PA ranges in \autoref{fig:shock} show the transit of the CME front as an evident Ly$\alpha$ profile broadening, mainly due to bulk flows along the LOS.

\subsection{WL channel}

\begin{figure}
    \centering
    \includegraphics[width=0.5 \textwidth]{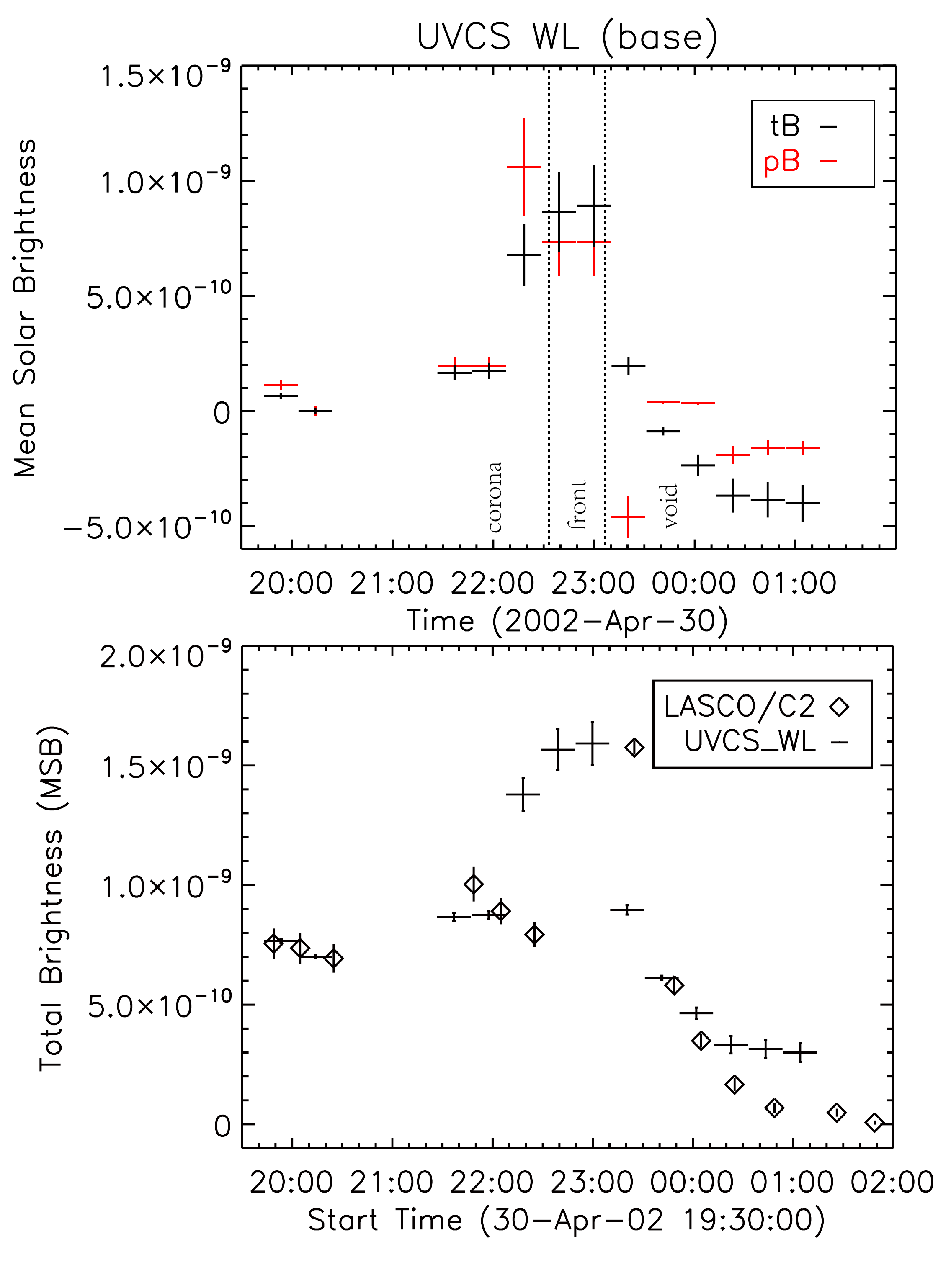}
    \caption{Top: the tB (black) and pB (red) intensities observed by the UVCS white-light (WL) channel after the pre-event intensity subtraction, at PA$=252^{\circ}$ corresponding to a latitude of 18$^\circ$ SW. The exposure time is around 20 minutes. Two vertical dotted lines are the same as that shown in the middle panel of \autoref{fig:shock}. Bottom: The UVCS tB intensities are re-calibrated (with a simple shift) based on the LASCO intensities with the minimal value subtraction. The UVCS WL error is 20\% of the photometric WL intensity and the LASCO WL error is about 3\% \citep{Frazin2002}. }
    \label{fig:calibration}
\end{figure}

This event was also observed by the WL channel of UVCS with $14\times14~\rm arcsec^2$ spatial field, which can provide the polarized (pB) and total (tB) brightness observations of the CME at the centre of the instantaneous UVCS FOV ($h = $2.45 $\rm R_{\odot}$ and PA$=252^{\circ}$, corresponding to 18$^\circ$ SW) with a 20-minute exposure time. Data acquired with the UVCS WL channel have been employed in the past for instance to constrain the radial dependence of electron density in the solar corona \citep{Miralles2001} or to study coronal density fluctuations and compressional waves \citep{Ofman1997,Ofman2000}. However, this is the first time for a CME to be analyzed combining observations from the UV and WL channels of UVCS instrument. The tB and pB WL intensities have been calibrated here by using the latest version (DAS 51) of the standard UVCS data analysis software distributed within SolarSoftware. The UVCS radiometric calibration error in WL is about 10\% given by \citet{Romoli2002} and about 7 \% given by \citet{Frazin2002}. UVCS and LASCO radiometric calibration differences are on the order of 20\%, with pB measured by LASCO being systematically larger by 20\% than that measured by UVCS \citep{Frazin2002}. Considering that the 20\% difference is more than the sum of uncertainties declared by the UVCS and LASCO WL data, we assume this 20\% as the final radiometric calibration uncertainty of UVCS WL channel, while the LASCO radiometric calibration error in WL is about 3\% \citep{Frazin2002}.
Nevertheless, despite the calibration activities of UVCS WL channel performed in the past \citep[e.g.][]{Romoli2002}, and systematic disagreements between UVCS and LASCO WL calibrated intensities were already reported by previous authors \citep[e.g.][]{Frazin2002}, a final correction for this disagreement was never implemented in the UVCS calibration software. Hence in this work, the UVCS WL intensities after minimal intensity background subtraction have been compared and ``re-calibrated" based on the LASCO intensities with 24hr-minimal background subtraction at the same position. In particular, in order to make the observations from the UVCS WL channel and from LASCO comparable, we added a constant shift by $3 \times10^{-10}$ mean solar brightness (MSB) to the UVCS WL intensities. This shift allows us to measure the same tB before the CME with the two instruments, as shown in \autoref{fig:calibration} (bottom panel). 

After this UVCS and LASCO WL inter-calibration, the WL intensities observed by UVCS during the CME can be considered as complementary data for the LASCO observations in the WL band-pass, because LASCO did not acquire polarized exposures during this event. In \autoref{fig:calibration} (top), red and black lines show the pB and tB data (respectively) at the centre of the UVCS slit, after the subtraction of the coronal background pre-CME intensity. The horizontal extension of each line is 20 min, corresponding to the WL channel exposure time. In \autoref{fig:calibration} (top), it is obvious that enhancements of the tB and pB intensities appeared between 22:10 UT and 22:30 UT, implying the possible arrival of the CME front passing at that time through the center of the UVCS slit position. Hence, these data provide us a possible timing for the arrival of the CME front in the UVCS FOV, that will help us in the interpretation of the UV intensity evolution. What's more, the eruption time of the CME source region must be earlier than 22:30 UT.% and also earlier than the predicted time (22:34 UT) inferred from automatic LASCO CDAW catalog.

Moreover, once we have the tB and pB intensities emitted by the CME plasma, we can use the polarization-ratio technique \citep{Moran2004} to estimate the possible position angle of the CME with respect to the POS at the latitudinal direction covered by the UVCS slit center, although we have only the one-pixel observation. During the front transit (time interval between 22:10 UT and 23:10 UT), when the pBs are larger than the tBs the polarization-ratio technique cannot be applied. During a CME, this could be partly related also with the evolving fraction of stray-light that, being not fully removed after instrument calibration, provides additional spurious signals that are different from one instrument to the other, leading to different intensity variations. Hence, for this event we only use the data points acquired in two other time ranges (22:30 UT to 22:50 UT and 22:50 UT to 23:10 UT) to derive the position angle, when the CME front was just captured by the WL channel; meanwhile, we ignore the meaningless result where the pB is larger than the tB due to the large uncertainty from the radiometric calibration. From the measured polarization ratio pB/tB during this time interval, the derived position angle is from $0^{\circ}$ to $26^{\circ}$ away from the POS. Because of the $\pm z$ ambiguity in the polarization-ratio technique \citep[see][]{Moran2004}, the position of the CME plasma in principle could be located in front of or behind the POS. Nevertheless, because as discussed before the source region of the CME is possible located on or behind the visible hemisphere, we can only make sure that the CME position angle is no more than $\sim26^{\circ}$ away from the POS at this PA. Despite the uncertainties in the polarization ratio technique related with LOS integration effects \citep[see e.g.][]{Moran2004,DaiXH2014,Susino2016,Lu2017}, \citet{Bemporad2015} demonstrated with numerical simulations that the best performances of this technique occurs when the CME propagation angle is around 20$^\circ$ from the POS. The derived position angle will be employed in the subsequent analysis and interpretation of these data.
\begin{figure}
    \centering
    \includegraphics[width=0.4\textwidth]{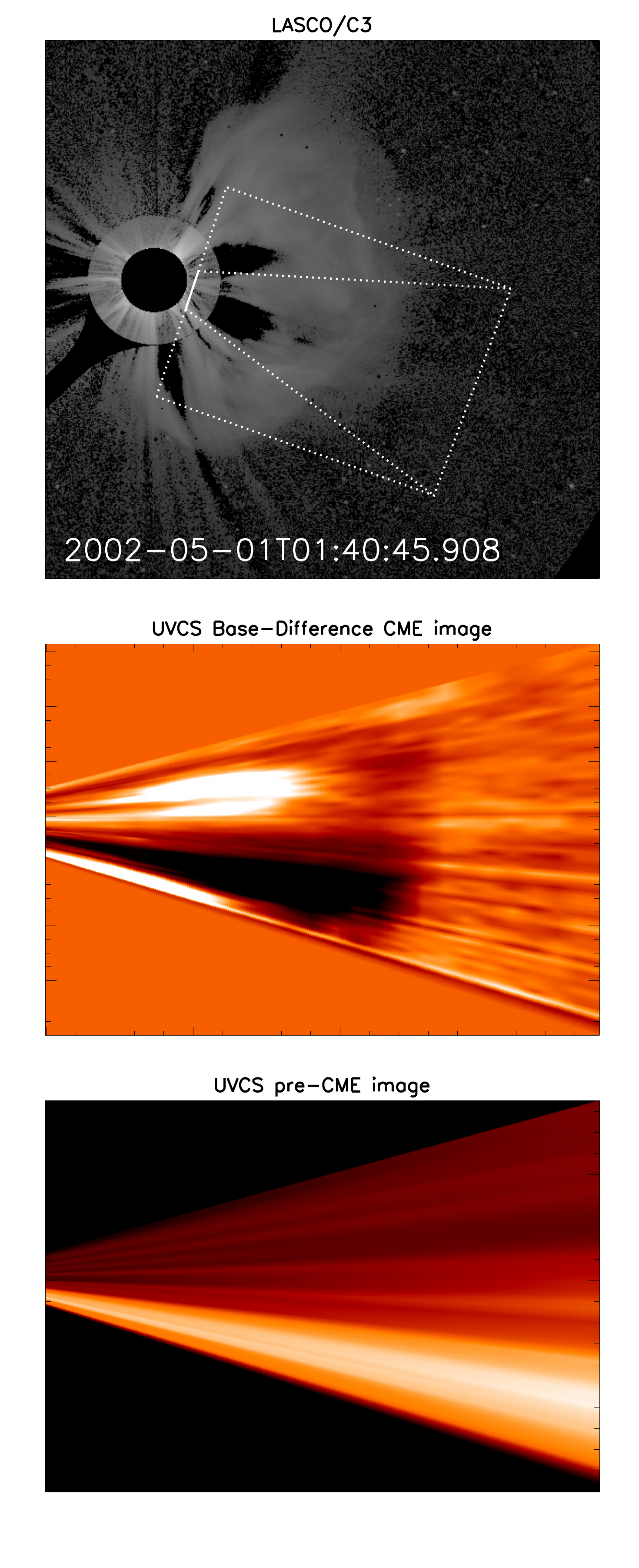}
    \caption{Top: LASCO/C3 base-difference image at 01:40 UT on 2002 May 1st. The solid white line represents the FOV of the UVCS spectrometer. The dotted box is the region of interest (ROI), whose position corresponds spatially to the middle panel. Middle: reconstructed CME image with a synthetic background subtraction derived from the UVCS H {\small I} Ly$\alpha$ intensity sequence. The edges on both sides of the UV CME image correspond to the dotted lines in the top panel. Bottom: a pre-CME coronal image made by linear interpolation of pre-CME intensity evolution observed as a function of time.}
    \label{fig:uvcs2d}
\end{figure}

\subsection{Comparison between the WL and UV channel}~\label{sec:com_uv_wl}
In order to directly compare the CME body appearance in WL and UV channels, we reconstructed a 2D UVCS (H {\small I} Ly$\alpha$) image by assuming uniform radial propagation speed of 1000 $\rm km~s^{-1}$ and isotropic tangential expansion. The resulting 2D Ly$\alpha$ image (\autoref{fig:uvcs2d}, middle panel) can be compared directly with the corresponding region marked by dotted lines in the LASCO/C3 image (\autoref{fig:uvcs2d}, top panel). In order to build the 2D Ly$\alpha$ image we also build a pre-CME coronal image, which is made by linear interpolation of pre-CME intensity evolution observed by UVCS as a function of time at different latitudes, by assuming that the observations relative to the corona stops at $\sim$21:47 UT just before the arrival of the CME front (\autoref{fig:uvcs2d}, bottom panel). The pre-CME coronal Ly$\alpha$ image shows a bright feature located southward, and this feature corresponds to a small streamer complex visible in the WL LASCO C2 images; this streamer is then compressed and deflected after the CME transit, and it is visible in the reconstructed 2D CME image.

The middle panel of \autoref{fig:uvcs2d} is thus obtained as a base-difference image by subtracting the pre-CME Ly$\alpha$ coronal image to the CME Ly$\alpha$ image. This Ly$\alpha$ CME image shows again that the ``dimming front" is the real CME front from a comparison with the evolution of WL intensity acquired at same times and same latitudes, and the CME core is only visible in the UV lines. 
%\begin{figure}
    %\centering
    %\includegraphics[width=0.9\textwidth]{lya_doppler_speed.eps}
    %\caption{Left: the evolution of Ly$\alpha$ line intensities after the subtraction of the average pre-CME intensities. Right: the evolution of the Doppler shift speed derived from the Ly$\alpha$ line profiles with the single Gaussian fit. The core and front of the CME are marked by black lines.}
    %\label{fig:doppler}
%\end{figure}

%%%%%%%%%%%%%%%%%%%%%%%%%%%%%%%%%%%%%%%%%%%%%%%%%%%
\section{Temperature Diagnosis with the Combination of the WL and UV Observations}
\label{sec:CME_tmperature}

\subsection{Formation of UV emission}
\label{sec:Formation}

The formation mechanism of many strong coronal lines in UV (such as H {\small I} Ly$\alpha$ and O {\small VI} doublet lines) is a combination of radiative and collisional excitations, followed by spontaneous emission. Combining the UV and WL observations, it is possible to estimate the electron temperatures of CMEs, if the electron density is derived from WL data by assuming a reasonable thickness of the CME along the LOS. In particular, the radiative component of the H {\small I} Ly$\alpha$ line, is due to the resonant scattering by coronal neutral hydrogen atom of the Ly$\alpha$ chromospheric emission, and can be expressed as
\begin{equation}
    \centering
    I_{rad}\approx\frac{b~h~B_{12}~\lambda_0}{4\pi}~\frac{\Omega_{\odot}}{4\pi}~F_D(v_{rad})~\int_{LOS}n_i dl,
    \label{eq:rad}
\end{equation}
where $b$ is branching ratio for radiative de-excitation, $h$ is the Plank constant, $B_{12}$ is the Einstein absorption coefficient for the H atom transition, $\lambda_0$ is the reference wavelength of the transition, $\Omega_{\odot}$ is the solid angle of the solar disk at the scattering location, $n_i$ is the neutral hydrogen number density. The term $F_D$ is the so-called Doppler dimming factor, which can be expressed as \citep{Noci1987}: 
\begin{equation}
    \centering
    F_D(v_{rad})={\int_0}^{\infty} I_{\odot}(\lambda-\delta\lambda)\Phi(\lambda-\lambda_0)d\lambda.
    \label{eq:dimming}
\end{equation}
The term $I_{\odot}(\lambda-\delta\lambda)$ is the intensity spectrum of incident chromospheric radiation, $\delta\lambda=\lambda_0 (v_{rad}/c)$ is the Doppler shift of the incident profile due to the radial velocity $v_{rad}$ of scattering atoms, $c$ is the light speed, and $\Phi(\lambda-\lambda_0)$ is the normalized coronal absorption profile along the direction of the incident radiation.

The approximate expression of the H {\small I} Ly$\alpha$ collisional component, due to the de-excitation of a coronal hydrogen atom previously excited by collision with a free electron, is given by
\begin{equation}
    \centering
    I_{col}\approx\frac{b}{4\pi}~q_{col}(T_e)~\int_{LOS}n_e~n_idl
    \label{eq:col}
\end{equation}
where $q_{col}(T_e)$ is the collisional excitation rate, a function of the electron temperature, $T_e$.

Notice that the above expressions have been approximated by extracting the quantities $\Omega_{\odot}$, $F_D(v_{rad})$, and $q_{col}(T_e)$ from the integral along the LOS, which means that we are considering the average values of these quantities along the LOS. This is a common approximation that can be applied to the analysis of UV emission from CMEs, whose extension of emitting region along the LOS is expected to be more limited with respect to the UV observations of large scale coronal structures such as streamers and coronal holes. Moreover, according to the method applied by \citet{Susino2016}, the equations \autoref{eq:rad} and \autoref{eq:col} can be further simplified to
\begin{equation}
    \centering
    I_{rad}\propto R(T_e)~F_D(v_{rad})~ \langle n_e \rangle ~ L,
    \label{eq:rad_sim}
\end{equation}
\begin{equation}
    \centering
    I_{col}\propto R(T_e)~q_{col}(T_e)~\langle n_e \rangle^2~L.
    \label{eq:col_sim}
\end{equation}
where the hydrogen number density is $n_i\approx0.83~A_{el}~R(T_e)~n_e$ \citep{Withbroe1982}, the factor 0.83 is the ratio of proton to electron densities by assuming a fully ionized plasma with 90\% of H and 10\% of He, $A_{el}$ is the elemental abundance relative to Hydrogen (1 in this case), $R(T_e)$ is the elemental ionization fraction as a function of the electron temperature $T_e$, $L$ is the thickness along the LOS of the plasma emitting region, and $\langle n_e \rangle$ is the average CME electron density along the LOS.

\subsection{Determination of CME densities}

The average electron number density $\langle n_e \rangle$ (units of $\rm cm^{-3}$) can be derived from the WL tB images \citep[see][]{Quemerais2002} by assuming a reasonable thickness $L$. In fact, $n_e$ depends on the column density $N_e$ (units of $\rm cm^{-2}$) as $N_e =\int_{LOS} n_e dl~\approx~\langle n_e \rangle~L $. Usually, for the analysis of CMEs, the electron density is determined from the WL images related to the CME after the subtraction of the pre-CME images. Nevertheless, for the analysis of this event the above method cannot be applied. The reason is that this work is aimed at determining the CME densities and temperatures by combining WL and UV intensities. To this end, the WL and UV intensities to be compared have to be emitted by the same column of plasma. Nevertheless, during the CME transit the Ly$\alpha$ intensity decreases in particular at the front due to the Doppler dimming effect and plasma heating \citep[see discussion by][]{Bemporad2018}, then the background intensity removal lead to the negative UV intensities, which cannot be used for the analysis \citep[see also][]{Susino2016}. Hence, in order to compare the WL and UV emission from the same plasma, the densities here have been derived without removal of WL coronal background. This is a very important peculiarity of the analysis described here, and will be a crucial point in the future analyses based on the combination of images acquired by UV Ly$\alpha$ and WL coronagraphs such as Metis and LST.

\begin{figure}
    \centering
    \includegraphics[width=0.8\textwidth]{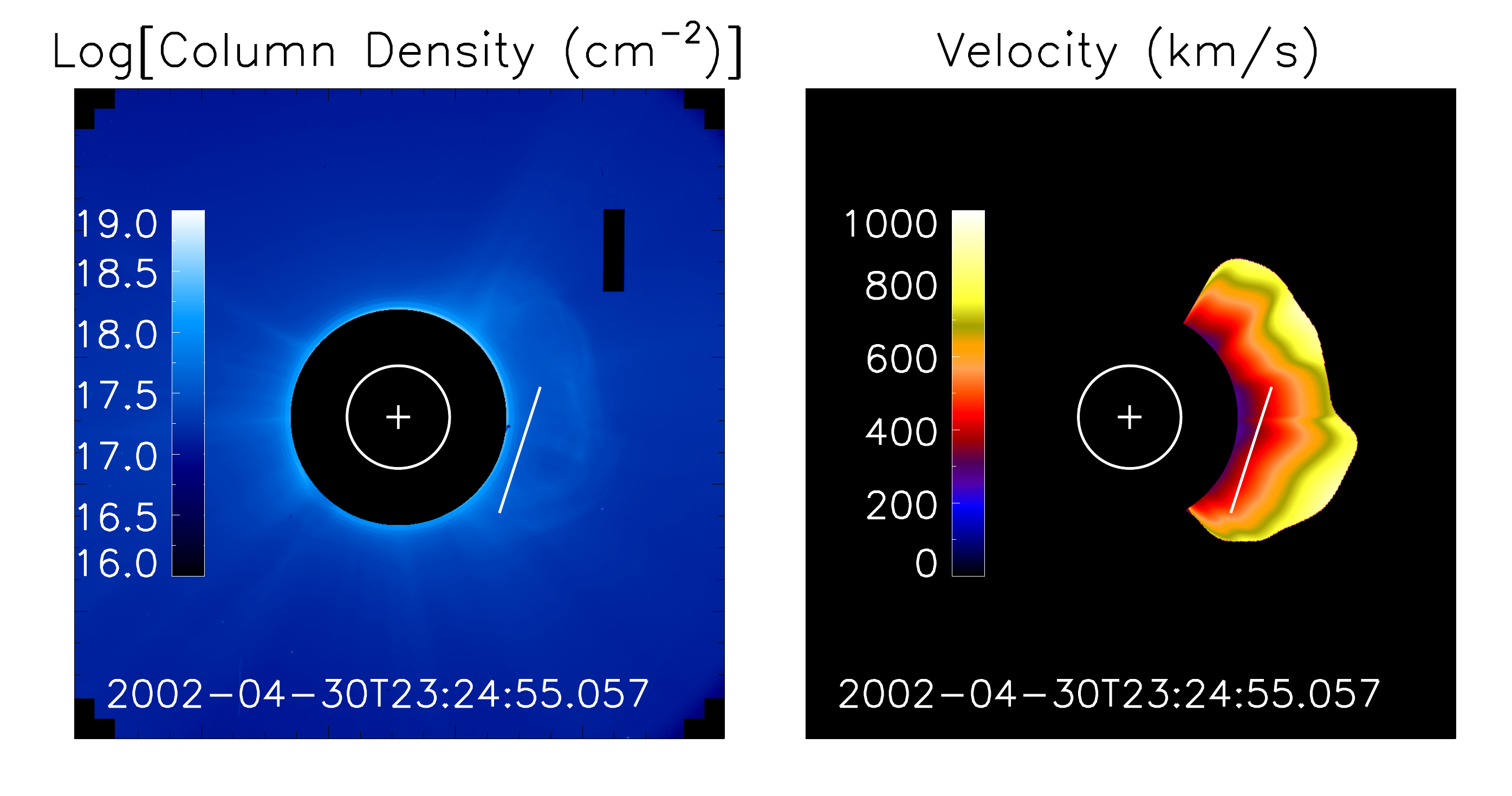}
    \caption{Left panel: 2D map of the electron column density $N_e$ derived from the LASCO/C2 tB image at 23:25 UT. Right panel: 2D map of the CME radial velocity derived from the sequence of WL images with a geometrical assumption (see text in \autoref{sec:speed}). White circles and plus signs denote the solar disk and its center; white solid lines represent the location of UVCS FOV.}
    \label{fig:cme2dspeed}
\end{figure}

\begin{figure}
    \centering
    \includegraphics[width=1.\textwidth]{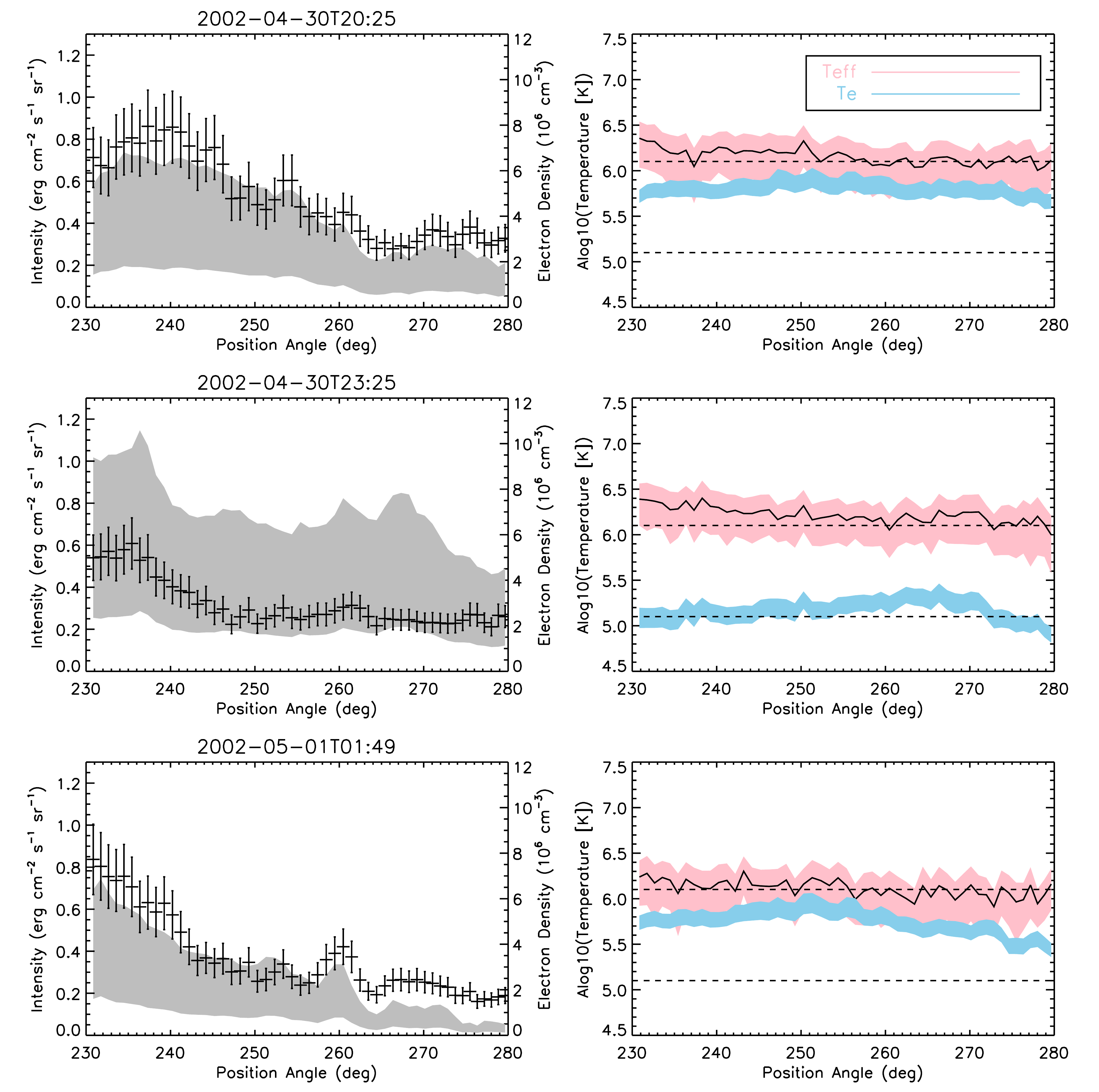}
    \caption{Left column: Evolution of electron densities (gray shaded region) and UVCS H {\small I} Ly$\alpha$ total intensities (plus symbols) as functions of the PA along the UVCS FOV before (top), during (middle), and after (bottom) the eruption of the CME. Electron densities, with the 3\% WL radiometric calibration errors, are obtained using two values of the thickness of the CME along the LOS, 0.25 $\rm R_{\odot}$ and 1 $\rm R_{\odot}$, and the uncertainty of the Ly$\alpha$ total intensity is 22\%. Right column: Corresponding evolution of hydrogen effective temperatures (pink shaded region) derived from the UVCS spectra with 1$\sigma$ uncertainty from a single Gaussian fit. The solid lines denote the average effective temperatures. The blue shaded regions represent the electron temperatures estimated from the combination of the WL and UV observations at different times, as functions of the PA along the UVCS FOV. The uncertainties of the electron temperatures are propagated from both the UVCS UV and LASCO WL radiometric calibrations, the measurements of the CME velocities, and two different thicknesses (0.25 and 1 $\rm R_{\odot}$) of the CME along the LOS, as well. Two dashed lines mark two fixed temperatures of $10^{5.1}$ and $10^{6.1}$ K. }
    \label{fig:tmperature}
\end{figure}

\autoref{fig:cme2dspeed} (left panel) shows a 2D map of the electron column density $N_e$ as derived without background subtraction from the LASCO/C2 tB image acquired at 23:25 UT (see also top middle panel of \autoref{fig:c23_img}). Because no polarized sequence was acquired by LASCO during this event, this map is derived from tB and not from pB, hence it includes also the spurious emissions from the photospheric light scattered by interplanetary dust (the F-corona, usually assumed to be negligible in pB images), and the emission from the coronal background, and will subsequently result in an overestimate of the real electron column density of the CME. In fact, the F-corona has been shown to be the brightest component of the continuum corona beyond about 2 $R_{\odot}$ \citep{VandeHulst1950,Koutchmy1985,Kimura1998}. Nevertheless, the F-corona is also exceptionally constant over the solar cycle \citep{Morgan2007}, thus, we use an existing model of the F-corona intensity and remove it from the continuum data, as is done by \citet{Dima2018}. As provided by \citet{VandeHulst1950}:
\begin{equation}
F(r)=14.86~r^{-7}+4.99~r^{-2.5},
\end{equation}
where $F(r)$ is the brightness of the F-corona in units of $10^{-8}$ MSB, as a function of the heliocentric distance $r$ measured in $R_{\odot}$. Notice also that, as pointed out by \citet{Ragot2003}, if a CME front owns a sufficiently strong normal component of the magnetic field, then the electromagnetic force could be strong enough to disperse the dusts beyond a few solar radii. In this kind of situation, the removal of the standard F-corona intensity with the above formula will result in an underestimation of the CME density.
Given these values of $N_e$, the average electron densities $\langle n_e \rangle$ have been derived from LASCO tB intensities with the subtraction of the F-corona brightness along the positions of the UVCS slit FOV and by assuming two different values of the CME thickness $L$ along the LOS: $L=0.25$ and $L=1$ $R_{\odot}$. These two assumptions of $L$ and the LASCO radiometric calibration error result in a gray shaded area of possible values of $ \langle n_e \rangle $ shown in \autoref{fig:tmperature} (left column) at three different times. In these panels, the plus symbols represent the H {\small I} Ly$\alpha$ total intensity observed by UVCS at the same times and locations, the error bars are 22\% of the Ly$\alpha$ total intensity \citep{Gardner2002}. In order to estimate the electron temperatures $T_e$ from a comparison between these curves, we need to calculate the Doppler dimming factors $F_D$, which are related with the radial CME velocity, the chromospheric intensity line profile, and the normalized coronal absorption profile. Hence, in what follows we derive an estimate of the radial CME speed at different latitudes (or PAs) along the UVCS slit and at different times. 

\subsection{Determination of CME internal radial speeds} \label{sec:speed}

Recently, a new technique to derive the 2D distribution of the radial CME speed was provided by \citet{Ying2019}. Nevertheless, this method is based on the cross-correlation analysis of WL coronagraphic images acquired by the STEREO/COR1 telescope with a time cadence of 5 minutes, and given the low cadence of LASCO images available for this event this technique cannot be applied here. Hence, we introduce an alternative (more simple) geometrical technique based on the analysis of LASCO WL image sequence; the same technique can be applied to any future coronagraphic observation of a CME acquired with low cadence. First, we convert the LASCO WL images from Cartesian to polar coordinates, and measure the distance $\Delta r$ between fronts of the CME at two different times (such as $T_n=$23:25 UT and $T_{n+1}=$23:49 UT on April 30). Then, the radial speed of the CME front $v_{POS}$, projected on the POS at different PAs, is derived at each latitude as $v_{POS}=\Delta r / \Delta T$. \citet{Feng2015a} analyzed the radial flow speed of a slow CME and found that its radial flow speed almost increased linearly with the distance along the radial direction. Thus, we assume here that the speed of the CME increases linearly along the radial direction, going from zero $\rm~km~s^{-1}$ at the solar surface up to the speed of the CME front. Finally, an example of the resulting 2D map of the CME radial speed as obtained at $\sim$23:25 UT is shown in \autoref{fig:cme2dspeed} (right panel). According to the above assumptions, the uncertainties in the location of CME front will lead to the uncertainties in the determination of CME speed, and further result in errors in the estimate of Doppler dimming factors. The uncertainty in the estimate of the CME front height is estimated to be 2 pixels ($\sim 0.024 ~\rm R_{\odot}$).

\begin{figure}
    \centering
    \includegraphics[width=0.8\textwidth]{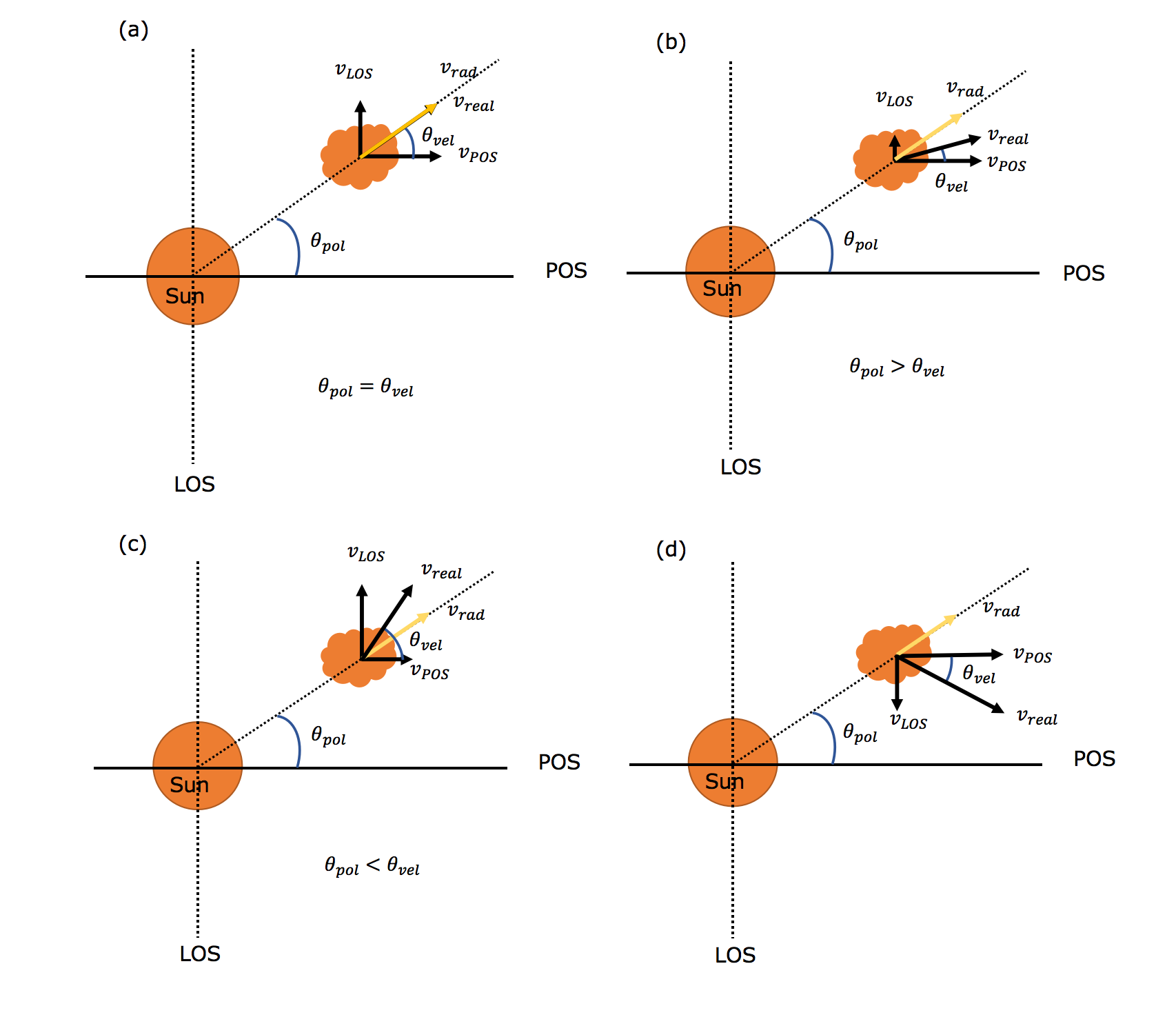}
    \caption{Possible different relationships between the real CME velocity $v_{real}$ and the radial velocity component $v_{rad}$ that needs to be measured to estimate the Ly$\alpha$ Doppler dimming factor $F_D(v_{rad})$. Panels (a), (b) and (c) show the plasma moves away from the POS, while the plasma moves towards the POS in panel (d). The measurable quantities are the plasma velocity projected on the POS ($v_{POS}$, from coronagraphic images), the velocity component along the line-of-sight ($v_{LOS}$, from Doppler shift of spectral lines), and the plasma position angle from the POS ($\theta_{pol}$, derived from the polarization-ratio technique). The angle $\theta_{vel}$ is the plasma propagation angle, which can be derived from the $v_{LOS}$ and $v_{POS}$ components.}
    \label{fig:sketch}
\end{figure}

It is important to point out here that the $v_{POS}$ measured with the above technique is in general an underestimate of the real radial speed $v_{rad}$, depending on the CME propagation direction with respect to the POS. The situation is illustrated in \autoref{fig:sketch}, as seen from an observer looking from above the ecliptic plane, with the POS represented by the solid black line, and the LOS represented by the dotted black line. In general, the real CME velocity $v_{real}$ is given by $v_{real}=\sqrt{v_{POS}^2+v_{LOS}^2}$, where for this event $v_{LOS}$ can be obtained directly from UVCS Doppler shift measurements, as it is shown in \autoref{fig:uvcs2d}, and $v_{POS}$ can be derived from coronagraphic images. If the CME has also a tangential velocity component, the measured real CME velocity $v_{real}$ is not aligned with the radial direction, but the radial projection of the CME speed is the quantity needed to estimate the values of Doppler dimming factor $F_D(v_{rad})$ to be applied to the analysis of the Ly$\alpha$ intensities. If the plasma moves away from the POS, when the plasma position angle $\theta_{pol}$ (derived from polarization ratio) is equal to its propagation angle $\theta_{vel}$, the plasma radial speed $v_{rad}$ coincides with the real velocity $v_{real}$, as shown in \autoref{fig:sketch} (panel a). In other conditions, as it is shown in \autoref{fig:sketch}, if $\theta_{vel} < \theta_{pol}$ (panel b) or $\theta_{vel} > \theta_{pol}$ (panel c), the $v_{POS}$ will be larger or smaller (respectively) than the plasma radial speed $v_{rad}$.
However, if the plasma moves towards the POS, the $v_{POS}$ will often be larger than the plasma radial speed $v_{rad}$. For this specific event, due to the uncertainty of the source region, we cannot be sure whether the CME plasma moves towards or away from the POS, although its LOS speed is blue-shifted in the front, while in the core it is mildly red-shifted in Ly$\alpha$ line and almost blue-shifted in the O {\small{VI}} lines. The maximal propagation angle $\theta_{vel}$ of this CME estimated by the $v_{LOS}$ (measured from the single-Gaussian fit, the top right panel of \autoref{fig:uvcs_img}) and the $v_{POS}$ is around $5^{\circ}$, while the CME position angle $\theta_{pol}$ is no more than $26^{\circ}$. For this CME, if we assume $\theta_{pol}=26^{\circ}$ and $\theta_{vel}=5^{\circ}$, the geometric relationships between the $v_{real}$, $v_{POS}$ and $v_{rad}$ can be established directly, according to the \autoref{fig:sketch} (panel b). Then, the relative difference (normalized to $v_{real}$) between $v_{rad}$ ($v_{POS}$) and $v_{real}$ is no more than 7\%. The smaller difference between $\theta_{pol}$ and $\theta_{vel}$ is , the closer value of $v_{rad}$ and $v_{real}$ will be. 

The combination of the $v_{LOS}$ obtained from the UVCS spectrometer and the $v_{POS}$ measured from the WL image sequence helped us to constrain the uncertainty of the radial velocity of the CME plasma. In the future, Metis and LST coronagraphs will provide us with higher temporal and spatial WL and UV observations, but without co-spatial slit-spectroscopic observation. Hence, it will be important to provide more accurate measurements of the CME radial speed and the CME propagation direction, that can be determined with the polarization-ratio method from single view-points, and 3D reconstructions \citep[e.g.][]{Feng2012a} from multi-point observations. However, this will be done bearing in mind that the radial speed estimated from the polarization-ratio method, which could provide us with plasma position angle ($\theta_{pol}$), will be over-estimated (or under-estimated), if $\theta_{pol} > \theta_{vel}$ (or $\theta_{pol} < \theta_{vel}$), as shown in \autoref{fig:sketch} (panels b and c), and often be over-estimated for the case as shown \autoref{fig:sketch} (d).

\subsection{Determination of CME temperatures}

In this work the UVCS data have been analyzed to derive both the electron temperatures (from Ly$\alpha$ intensities) and the kinetic temperatures (from Ly$\alpha$ line widths) during the transit of the CME. The line widths $\Delta\lambda_{1/e}$ of the H {\small I} Ly$\alpha$ line profiles were obtained from single-Gaussian fits and corrections for UVCS instrumental profile broadenings \citep{Kohl1999}. The $\Delta\lambda_{1/e}$ corrected for the instrumental profile broadening can be converted into effective temperatures, $T_{eff}$, as
\begin{equation}
    \centering
    \Delta\lambda_{1/e}=\frac{\lambda_0}{c}\sqrt{\frac{2k_B~T_{eff}}{m_H}},
\end{equation}
where $k_B$ is the Boltzmann constant, and $m_H$ is the hydrogen mass. As mentioned in \citet{Susino2016}, the observed profiles of the Ly$\alpha$ line are often broadened by non-thermal motions and by the plasma bulk velocity along the LOS. Thus the derived effective temperatures are only the upper limit of the real hydrogen kinetic temperatures. The uncertainty of the $T_{eff}$ comes from the Gaussian fit for the H {\small I} Ly$\alpha$ line with 1$\sigma$ error. %

\autoref{fig:tmperature} (right column) shows the effective temperatures in pink shades along the UVCS FOV at three different times starting from 20:25 UT on April 30 to 01:49 UT on May 1. At 23:25 UT, the core and void parts of the CME were captured by the UVCS slit, whose average effective temperature is around $10^{6.2}~\rm K$, a characteristic value of coronal conditions. The average effective temperatures observed at 20:25 UT of April 30 and 01:49 UT on the next day (before and after the eruption of the CME), respectively, seem a little bit lower than those measured at 23:25 UT. However, the uncertainties of the effective temperatures are so large that none of these changes are significant beyond the 1$\sigma$ level and thus are not distinguishable from data noise. The uniform $T_{eff}$ along the UVCS FOV has also been detected in \citet{Susino2016}. 

In order to measure the electron temperatures, the choice of the incident Ly$\alpha$ chromospheric radiation profile is essential; in this work we use the Ly$\alpha$ disk profile $I_\odot(\lambda)$ reported by \citet{Lemaire2002} and measured with SOHO/SUMER on 2001 August 22, very close to the date of this event. The atomic absorption profile $\Phi(\lambda-\lambda_0)$ is assumed to be a Gaussian profile with a $1/e$ line width equal to the Ly$\alpha$ line width measured by UVCS data (dotted-dashed lines in the top panels of \autoref{fig:shock}). Given the electron densities derived from WL, the expected total H {\small I} Ly$\alpha$ intensity can be estimated as a function of $T_e$, and then we can derive the real $T_e$ by comparing the H {\small I} Ly$\alpha$ intensity observed by UVCS with the expected total intensity, according to \autoref{eq:rad_sim} and \autoref{eq:col_sim} \citep{Susino2016}. 

The derived electron temperatures (blue shaded areas) are shown in \autoref{fig:tmperature} (right column) by assuming two different thicknesses $L=0.25$ and $1 \rm R_{\odot}$. Results show that the derived electron temperatures decrease at 23:25 UT, with respect to $T_e$ estimated at two other time intervals. What's more, at 23:25 UT, the average $T_e$ of $10^{5.2}$ K is about an order of magnitude lower than the average $T_{eff}$ of $10^{6.2}$ K, a different result with respect to the relationship between the $T_{eff}$ and $T_e$ shown in the work of \citet{Susino2016}. The uncertainty of $T_e$ comes from the uncertainties on the thickness along the LOS, the measurement of the CME speed, as well as the errors from both the UVCS UV and LASCO WL radiometric calibrations.

\begin{figure}
    \centering
    \includegraphics[width=1.\textwidth]{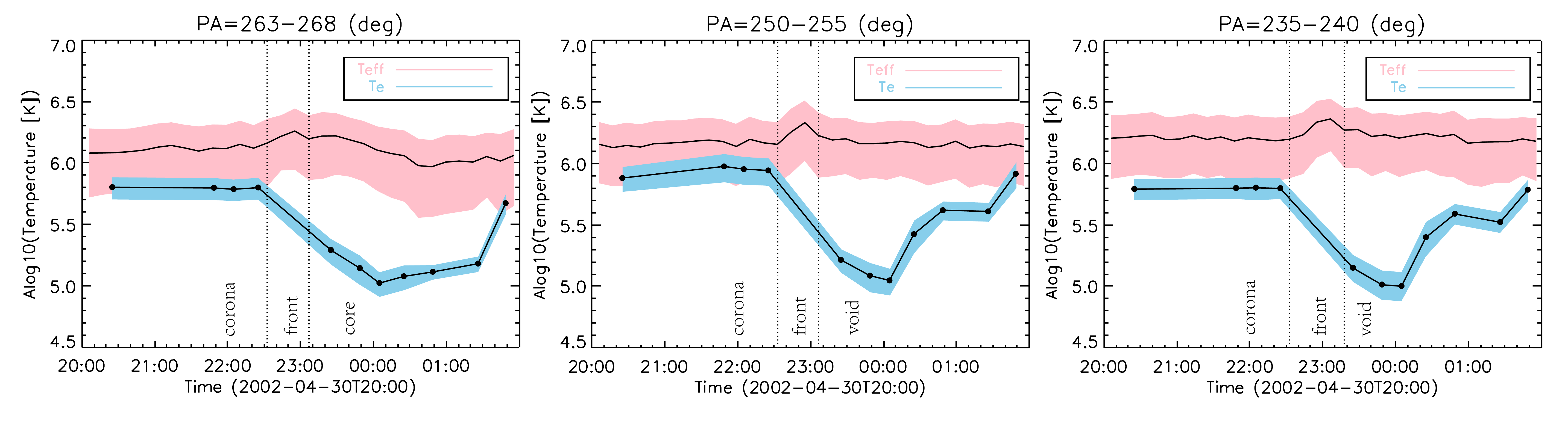}
    \caption{Time evolution of the electron and effective temperatures of the CME plasma averaged from different PA ranges (corresponding to those shown in \autoref{fig:shock}). The transits of different CME parts are shown by vertical dotted lines. Black dots denote times when WL observations are available. The blue shadow area marks the errors of the electron temperatures propagated from the measurements of the CME velocities, different LOS thicknesses, as well as uncertainties from both the UVCS UV and LASCO WL radiometric calibrations. The pink shade area shows the uncertainty of the effective temperatures of the CME derived from the single-Gaussian fitting with 1$\sigma$ uncertainty. }
    \label{fig:temp_tim}
\end{figure}

%\autoref{fig:temp_tim} (top panel) shows the latitudinal distribution of the \textbf{comparison} between CME electron temperatures measured along the UVCS FOV at 23:49 UT and 23:25 UT. These two times correspond to the crossing of the CME void region located Southward with respect to the core, as it is shown in \autoref{fig:uvcs_img} and \autoref{fig:uvcs2d}. \textbf{The core of the CME almost shows a ``cooling" process, suggesting that the plasma cooling due to the possible adiabatic expansion of the CME is stronger than any heating process. Further confirmation of the CME adiabatic expansion will be given below.}
The time evolution of CME plasma electron and effective temperatures is shown in \autoref{fig:temp_tim}. $T_e$ and $T_{eff}$ values are averaged with $5^{\circ}$, and the PA ranges correspond to the ranges shown in \autoref{fig:shock}. According to the evolution of the total H {\small I} Ly$\alpha$ intensity in \autoref{fig:shock} (left panel), the whole process can be divided into three phases as well: the coronal background, the CME front, and the CME core. At the other two PAs, the UVCS FOV only captures the CME void after the transit of the CME front.
Since only a small portion of the corresponding WL observations, the CME core and void are the only parts with complete observations, while the CME front was not captured by the LASCO observation in this event as it passed through the UVCS FOV. The lowest $T_e$ of the CME core and void even goes down to $10^{5}$ K. Interestingly, although the electron temperatures change dramatically, the hydrogen effective temperatures, in contrast, keep almost uniform, despite the transit of the CME (\autoref{fig:temp_tim}). Actually, many previous UVCS observations have found that the H {\small I} Ly$\alpha$ profiles are often narrower in the CME core, and the corresponding proton temperatures are also in the order of $10^{5}$ K \citep{Kohl2006}. Considering that the UVCS slit was centered at a projected altitude of 2.45 $\rm R_{\odot}$, at this large altitude the electron temperature is around $9\times10^5$ K in a coronal streamer \citep{Gibson1999} and as low as $7\times10^4$ K in a coronal hole \citep{Guhathakurta1999}, and the electron temperatures we derived fall in this interval.

It is obvious that the evolution of the electron temperature is similar between the CME core and void during 23:25 UT on April 30 and 00:04 UT of next day, which implies that the CME core and void might suffer the similar cooling process, such as the expansion cooling. In order to estimate the possible plasma cooling due to the CME expansion alone (neglecting other cooling processes such as conduction and radiative cooling), we estimated the volume of the CME by assuming a CME shape of a spherical sector \citep{Bein2013} at 23:25 UT and 23:49 UT, when the core and void of the CME are undergoing the cooling process. Setting the volume and electron temperature as $V_1$ and $T_{e1}$ at 23:25 UT, and $V_2$ and $T_{e2}$ at 23:49 UT, we verified that $T_{e1}~V_1^{\gamma-1} \approx T_{e2}~V_2^{\gamma-1}$ when $\gamma=4/3$ as expected for cooling due to expansion. The electron temperatures ($T_{e1}$ and $T_{e2}$) are obtained at three PA ranges as shown in \autoref{fig:temp_tim}. We should notice that the $\gamma=4/3$ is less than the typical adiabatic index of $\gamma=5/3$. \citet{Wang2009} have proposed a generic self-similar flux rope model to investigate the thermodynamic process and expansion of
CMEs. In this model, they discovered that the polytropic index value of $4/3$ is a critical point to judge whether with increasing distance the absolute value of Lorentz force decreases slower or faster than that of thermal pressure force. After the analysis of a real CME, the authors found that the polytropic index of the CME increased from 1.24 to 1.35 and subsequently slowed down to 1.336, which is close to the value of $4/3$. The polytropic index less than $5/3$ means the existence of heating processes in the CME core and void \citep{Wang2009}, but the cooling due to the CME expansion could dominate in the core and void, resulting in the observed decrease of the electron temperature for our event. %What's more, the occurrence of the electron temperature decrease, when the CME core transits through the UVCS FOV, might also be a signature of cold plasma in the erupting prominence embedded in the CME core.

Subsequently, the electron temperatures of the CME void (\autoref{fig:temp_tim}, middle and right panels) started to increase after 00:04 UT of the next day, while those of the CME core evolved slowly between 00:04 UT and 01:30 UT and rose obviously after 01:30 UT. The difference evolution of the electron temperature between the CME core and void might be owing to the existence of the prominence segments in the core preventing the temperature rise of the plasma at that time. Temperature enhancements of the CME void could be due to many different processes, such as the conversion of magnetic energy to thermal energy \citep{Bemporad2007}, the heating from the corona background as well as the different plasma conditions captured by the UVCS FOV \citep{Susino2016}. \citet{Glesener2013} found that there existed non-thermal plasma in the CME core contributing to the heating of the CME. Thus, the much higher effective temperatures in the CME core and void might indicate that the possibly lower kinetic temperatures of the CME core and void are counter-balanced by higher non-thermal plasma motions. The bright UV core in the Ly$\alpha$ line (in \autoref{fig:uvcs_img}) might be explained by the low ionization temperatures of hydrogen atoms, leading to higher abundances and thus observed intensities.

\begin{figure}
    \centering
    \includegraphics[width=0.8\textwidth]{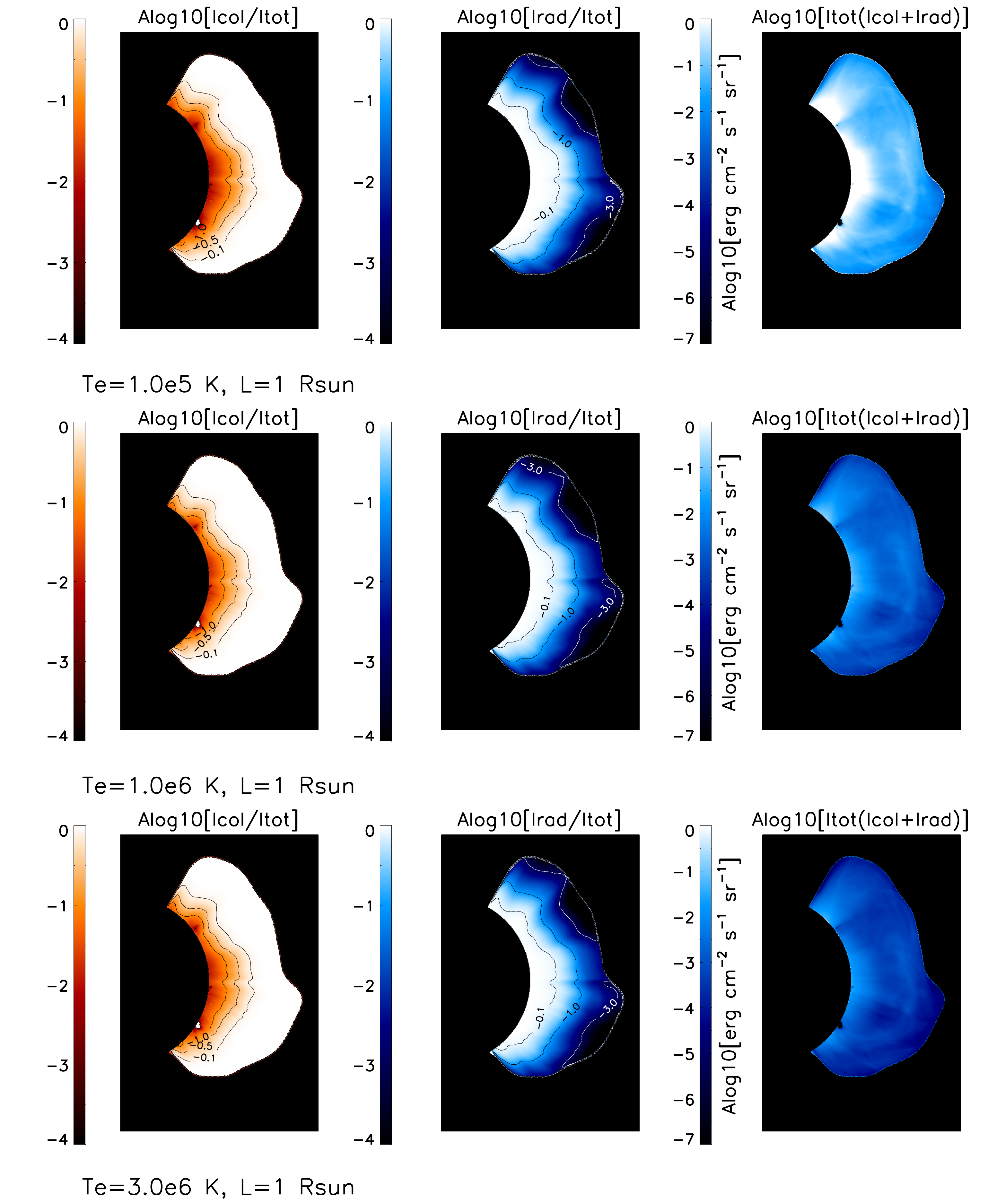}
    \caption{Synthetic UV images of the H {\small I} Ly$\alpha$ intensity with different assumptions of the electron temperature ($1\times10^5$ K, $1\times 10^6$ K, and $3\times10^6$ K). The corresponding electron density is derived from the LASCO WL image at 23:25 UT with 24hr-minimal background subtraction. The assumption of the $L$ along the LOS is equal to 1 solar radius. From left to right columns, these images show the collisional component (red) to the total intensity, radiative component (blue) to total intensity and the total intensity (collisional plus radiative components) in logarithm scales, respectively. Contour lines are plotted on the left and middle columns.}
    \label{fig:recon}
\end{figure}

\citet{Bemporad2018} synthesized the UV Ly$\alpha$ images from MHD simulations; in this work we also reconstruct synthetic UV images in the H {\small I} Ly$\alpha$, but we perform this task based on real observations (instead of synthetic data) and by using the simplified equations \autoref{eq:rad_sim} and \autoref{eq:col_sim}. Resulting images are shown in \autoref{fig:recon}. Due to the lack of measurements of the electron temperature distribution within the whole CME body, we assume here three electron temperatures ($1\times 10^{5}$ K, $1\times 10^{6}$ K, $3\times 10^{6}$ K). A value of $1\times 10^{5}$ K is the typical electron temperature of the CME core measured in our event, while the other two values have been reported by \citet{Susino2016}. The electron column density from WL images has been converted to the average electron density with 24hr-minimal background subtraction and with an assumption of $L=1 \rm R_{\odot}$. Thus, the reconstructed UV intensities represents only the contribution from the CME plasma without emissions from coronal foreground and background. To estimate the radiative component, the Doppler dimming factors have been constrained by the 2D radial speed of the CME. In \autoref{fig:recon}, the three columns from left to right represent the collisional (red) and radiative (blue) components with respect to the total intensity, respectively, as well as the H {\small I} Ly$\alpha$ total intensity. Although these results are based on very simple assumptions, they still show some very interesting consequence. Concerning the synthetic total intensity, the front emissions of the CME mainly come from the collisional component due to the fast speed, while the radiative component dominates in the inner part of the CME. \autoref{fig:recon} (right column panels) also show that, as expected, the higher the electron temperature is, the lower the total Ly$\alpha$ intensity will be. Similar analyses might help to constrain the 2D distribution of average electron temperatures in CME bodies, with a comparison between the reconstructed and the future real observations of CMEs in UV Ly$\alpha$ with the Metis and LST instruments.

%%%%%%%%%%%%%%%%%%%%%%%%%%%%%%%%%%%%%%%%%%%%%%%%%
\section{Discussions and Conclusions}
\label{sec:discussion_conclusion}
In this work, we have analyzed a fast CME with a driven shock to extensively develop and test the diagnostics that will be applied to derive plasma parameters of CMEs using future multi-wavelength observations from the Metis and the LST coronagraphs. Due to the lack of sufficient observations, it was hard to firmly identify the source region of this eruption. The combination of the SOHO/EIT 195 {\AA} and TRACE 195 {\AA} images, LASCO C2 images, together with magnetic field lines extrapolated with PFSS method suggests that the source region could be located on the visible hemisphere and associated with a jet eruption having time and direction consistent with the CME. However, we cannot ensure whether this small jet could provide enough twist to the overlying loop system to destablize it and result or not in the observed fast partial-halo CME.

Combining the SOHO/LASCO WL images and the UVCS spectra, we have derived physical parameters of the CME plasma, including its 3D propagation direction, 2D radial speed map, as well as effective and electron temperatures. The observations of a CME acquired by the UVCS WL channel have been analyzed here for the first time. The main conclusions of this work are as follows:
%According to the diagnostic methods \citep{Susino2016,Bemporad2018}, which will be applied to coronagraphic observations of CMEs delivered by the Metis instrument onboard the next ESA-Solar Orbiter mission and the LST onboard ASO-S mission, we combine the WL data from LASCO/C2 and UV H {\small I} Ly$\alpha$ line (1216 {\AA}) from the UVCS spectrometer to acquire the information of the CME plasma temperatures, including kinetic and electron temperatures. 

1. The CME position angle with respect to the POS, as derived from the polarization-ratio technique applied to the UVCS WL channel data, is no more than $26^{\circ}$. This angle was compared here with the CME propagation angle of about $5^\circ$, as derived from a kinematical analysis based on the WL images and the measured Doppler shift of spectral lines in UV. This result reminds that in general CMEs are not simply expanding isotropically along the radial direction, hence the analysis of single view-point projected images provided by a single coronagraph are in general not sufficient to fully characterize the CME kinematic, and slit-spectroscopic observations or multi view-point coronagraphic observations are also needed.
%
%2. The CME-driven shock possible crossed the UVCS slit between 22:10 and 22:30 UT. The transit of the shock region might contribute to the enhancement of tB and pB values observed by the UVCS/WL channel, and could result in the broadening of H {\small I} Ly$\alpha$ line profile as observed by the UVCS spectrometer. Subsequently, the shock front propagated into the FOV of the LASCO C2 and C3 coronagraphs resulting in the classical faint front preceding the CME leading edge.

2. This CME appears very different when observed in the WL and UV band-passes. The CME shows a dimming front and a bright core in the UV intensities, while it has a typical bright front but no visible core in the WL images acquired at same times and same latitudes. One of possible reasons for the absence of a clear core in the WL images might be the orientation of the neutral line (hence the CME axis) with respect to the LOS direction. As showed by \citet{Bemporad2018}, the different appearance of the CME between the WL and UV Ly$\alpha$ intensities is mainly due to the Doppler dimming effect and the distribution of plasma temperatures. In this event, the CME core might own a lower electron temperature and higher density thus being bright in UV, while the dark dimming front in the UV image might be primarily due to the fast radial speeds (low values of the Doppler dimming factors) and higher plasma temperatures.

3. The UV Ly$\alpha$ intensities due to the radiative excitation are always affected by the Doppler dimming effect in CMEs, resulting from the Doppler shift between the chromospheric emission and the coronal absorption profiles. CMEs are usually characterized by the presence of significant inhomogeneities not only in the plasma density, but also in the distribution of plasma velocity, because different parts of the CME are propagating outward with different radial and latitudinal speeds. Thus, the one-dimensional average velocity of the CME usually determined from the displacement of CME front will lead to large errors in the estimate of the Doppler dimming factors by assuming the same velocity across the CME body. Hence, we obtain here the 2D distribution of the radial speed map by measuring the front speed at different latitudes and by assuming that the radial speed increases linearly at each latitude along the radial orientation. In the future, the radial velocities of CMEs, observed by new instruments with higher temporal and spatial resolutions (such as Metis and the LST), will be better determined by the cross-correlation method recently provided by \citet{Ying2019}. Subsequently, the WL tB images with the subtraction of the F-corona intensity are converted to the electron densities along the UVCS FOV, by assuming two thicknesses (0.25 and 1 $\rm R_{\odot}$) of the CME along the LOS. Then combining the derived electron densities with the observed UV Ly$\alpha$ total intensities, we estimated the electron temperatures of the CME (and background corona) with diagnostic methods described by \citet{Susino2016} and \citet{Bemporad2018}. 

4. The evolution of the electron and effective temperatures before, during, and after the transit of the CME along the UVCS FOV (\autoref{fig:tmperature}) was explored. At 23:25 UT, the CME core and void were observed by the UVCS slit, and we found that the CME plasma might own a uniform effective temperature with an average value of $\sim 10^{6.2}$ K along the UVCS FOV, while the average electron temperature is one order of magnitude ($\sim 10^{5.2}$ K) lower than the plasma effective temperature. %\textbf{The electron temperature comparison between 23:49 UT and 23:25 UT shows that the core of the CME is characterized by plasma cooling.} 

5. At PA$=263^{\circ}-268^{\circ}$, the UVCS slit captures different parts of this eruptive event, including the pre-event coronal background, the CME front, as well as the CME core,while at the other two PAs UVCS FOV only observe the CME void after the transit of the CME front. The lowest electron temperatures of the CME core and void we measured are around $10^{5}$ K. The observed cold plasma in the core might result from possible prominence segments and the plasma cooling due to the CME expansion occurring at a rate higher than any possible plasma heating process. The electron temperature decreases of the CME void might be due to the CME expansion. On the other hand, the subsequent electron temperature increase in the CME void observed after 00:04 UT in the next day might imply the conversion of magnetic energy into thermal energy \citep{Bemporad2007}, and the heating from the corona background as well as the different plasma conditions captured by the UVCS FOV \citep{Susino2016}. The heterogeneity between the effective and electron temperatures might indicate that in the CME core and void the potentially lower kinetic temperatures are counter-balanced by higher non-thermal plasma motions. What's more, we have reconstructed synthetic UV images in the H {\small I} Ly$\alpha$ line based on the WL observations by assuming different temperatures to show how future UV observations of CMEs provided by the Metis and LST instruments will be affected by different distributions of plasma temperatures in the bodies of CMEs.
%, the electron temperature decreased first and then nearly returned to the previous electron temperature in this region, which might imply the adiabatic expansion of the CME. What's more, the kinetic temperature is obviously higher than the electron temperature during the transit of the CME, which is different with the results of Susino et al. (2016), while in that work, the magnitude of the electron temperature and the proton temperature of three slow CME are comparative.Given the absence of a clear prominence embedded in this CME,

We remind here that, as pointed out by \citet{Bemporad2018}, the electron temperatures derived with methods applied here are likely underestimated (by $\sim$40\%-50\%) taking into account the LOS averages. The investigators suggested to use the fully collisional excitation (FCE) assumption to derive the electron temperatures of the CME core, while for the remainder of the CME the radiative and collisional excitation (RCE) approximation works much better. In our work, we only use the RCE approximation and not the FCE approximation to estimate the CME electron temperature. Although the underestimate of the electron temperature is not negligible, the obvious descent and ascent of the electron temperature of the CME core are still reliable.
 %In this work, the modified radial velocity of the CME is obtained by combining the Doppler shift speed from the UVCS spectrometer and the projected radial speed on the POS from the WL images. In the fact, the expansion of the CME in all directions might result in the underestimation of the Doppler shift speed along the LOS, due to the counteraction of speeds in opposite directions. When the position angle ($19^{\circ}$) of the CME front has been estimated, we also could obtain its modified radial velocity. For this event, the difference between these two kinds of modified radial velocity is no more than 5\%. Thus, the Doppler shift speed did not affect a lot by the CME expansion in this event.

The combination of the WL and UV observations has shown its power to reveal new information of physical processes, including not only shocks \citep{Bemporad2010}, but also CMEs \citep{Ciaravella2005,Bemporad2007,Bemporad2018,Susino2016} and eruptive prominences \citep{Heinzel2016,Susino2018}. In the future, the simultaneous observation of the WL and UV (H {\small I} Ly$\alpha$ line) provided by the LST and Metis instruments would be helpful to estimate the physical parameters of solar eruptions and understand the physical mechanisms responsible for the plasma evolution during the CME expansion phases.

This work demonstrates that the combination of WL and UV Ly$\alpha$ intensity evolution acquired at the same time during the transit of a CME is very important to allow the identification of different CME features (i.e. front, void, core) in UV data. In fact, these features are usually defined and identified form WL images, but the CME appears totally different in UV, hence without the WL counterpart it would be much more complicated to identify the location of these features in UV images. Hence, this work shows the potentiality of future WL$+$UV coronagraphic observations of CMEs, as will be provided by future coronagraphs.

It is also very important to point out that the electron temperatures of CMEs are usually estimated under the assumption of the ionization equilibrium. However, the corona is out of equilibrium in general already at the heliocentric distances seen by LASCO/C2, even in a stationary state. With theoretical and observational analysis, \citet{Landi2012} and \citet{Boe2018} reported that charge state compositions of heavy ions will freeze-in within 2 $R_{\odot}$ or so, and so the coronal plasma is leaving equilibrium even before 2 $R_{\odot}$. The freeze-in distances of the lighter elements stop evolving much earlier than that of the heavier elements \citep{Landi2012}. On the other hand, the freeze-in distances of the charge state compositions could be raised, if there exist bound structures \citep[e.g. coronal loops and streamers,][]{Boe2018}. Recently, with the MHD simulations, \citet{Pagano2020}  have reconstructed the ionization state of hydrogen atoms in a simulated CME, and found that there is a significant non-equilibrium ionization effect in the CME front, while in the CME core the equilibrium ionization assumption is still valid. Thus, the assumption of the ionization equilibrium in the data inversion would lead to an important plasma temperature underestimation in the CME front. 

Obviously, there are limitations of estimating electron temperature of CMEs only through observations provided by UV Ly$\alpha$ lines and WL intensities. Although UVCS observations have reported strong Ly$\alpha$ emissions from regions in the inner corona where the electron temperatures are up to $1.5\times10^{6}$ K \citep{Kohl2006}, due to the low ionization temperatures of the hydrogen atoms, the estimated electron temperatures are only valid for the plasma along the LOS that will be emitting Ly$\alpha$  emissions in the first place, which will never be the hottest regions. A more complete inference of the electron temperature requires additional lines that span the temperature range of the corona \citep{Raymond1997}. In the future, the observations of the visible emission lines, provided by the ASPIICS \citep{Lamy2010} aboard \textit{PROBA-3} mission and the Visible Emission Line Coronagraph \citep[VELC,][]{Prasad2017} aboard \textit{Indian Aditya-L1} mission, could be useful to derive the electron temperatures of the hotter parts of CMEs. Both these two coronagraphs will observe the low corona in the Fe XIV (5303 \AA) emission line, whose temperature is $\log T>$6. Moreover, the Cryogenic Near Infra-Red Spectro-Polarimeter (Cryo-NIRSP) filter of \textit{the Daniel K. Inouye Solar Telescope} \citep[DKIST,][]{Tritschler2015} on the ground owns many infrared lines (1- 5 micron), whose temperature sensitivity is up to $\sim$2 MK. The observations of the DKIST/ Cryo-NIRSP will also provide data for better constraining the electron temperature and electron density.

\acknowledgements 
 The authors acknowledge S. Giordano for his assistance on the UVCS CME data catalog. We also thank the anonymous referee for suggestions and comments that helped us significantly improve and clarify this paper. SOHO is a mission of international cooperation between ESA and NASA. This work is supported by NSFC (grant Nos. U1731241, 11921003, 11973012), CAS Strategic Pioneer Program on Space Science (grant Nos., XDA15052200, XDA15320103, and XDA15320301), the mobility program (M-0068) of the Sino-German Science Center and National Key R\&D Program of China (2018YFA0404200). L.F. also acknowledges the Youth Innovation Promotion Association for financial support.
\bibliography{refs}

\begin{thebibliography}{}
\expandafter\ifx\csname natexlab\endcsname\relax\def\natexlab#1{#1}\fi

\bibitem[{{Akmal} {et~al.}(2001){Akmal}, {Raymond}, {Vourlidas}, {Thompson},
  {Ciaravella}, {Ko}, {Uzzo}, \& {Wu}}]{Akmal2001}
{Akmal}, A., {Raymond}, J.~C., {Vourlidas}, A., {et~al.} 2001, \apj, 553, 922

\bibitem[{{Antonucci} {et~al.}(2017){Antonucci}, {Andretta}, {Cesare},
  {Ciaravella}, {Doschek}, {Fineschi}, {Giordano}, {Lamy}, {Moses}, {Naletto},
  {Newmark}, {Poletto}, {Romoli}, {Solanki}, {Spadaro}, {Teriaca}, \&
  {Zangrilli}}]{Antonucci2017}
{Antonucci}, E., {Andretta}, V., {Cesare}, S., {et~al.} 2017, in Society of
  Photo-Optical Instrumentation Engineers (SPIE) Conference Series, Vol. 10566,
  Society of Photo-Optical Instrumentation Engineers (SPIE) Conference Series,
  105660L

\bibitem[{{Beckers} \& {Chipman}(1974)}]{Beckers1974}
{Beckers}, J.~M., \& {Chipman}, E. 1974, \solphys, 34, 151

\bibitem[{{Bein} {et~al.}(2013){Bein}, {Temmer}, {Vourlidas}, {Veronig}, \&
  {Utz}}]{Bein2013}
{Bein}, B.~M., {Temmer}, M., {Vourlidas}, A., {Veronig}, A.~M., \& {Utz}, D.
  2013, \apj, 768, 31

\bibitem[{{Bemporad}(2008)}]{Bemporad2008}
{Bemporad}, A. 2008, \apj, 689, 572

\bibitem[{{Bemporad} \& {Mancuso}(2010)}]{Bemporad2010}
{Bemporad}, A., \& {Mancuso}, S. 2010, \apj, 720, 130

\bibitem[{{Bemporad} \& {Pagano}(2015)}]{Bemporad2015}
{Bemporad}, A., \& {Pagano}, P. 2015, \aap, 576, A93

\bibitem[{{Bemporad} {et~al.}(2018){Bemporad}, {Pagano}, \&
  {Giordano}}]{Bemporad2018}
{Bemporad}, A., {Pagano}, P., \& {Giordano}, S. 2018, \aap, 619, A25

\bibitem[{{Bemporad} {et~al.}(2007){Bemporad}, {Raymond}, {Poletto}, \&
  {Romoli}}]{Bemporad2007}
{Bemporad}, A., {Raymond}, J., {Poletto}, G., \& {Romoli}, M. 2007, \apj, 655,
  576

\bibitem[{{Bemporad} {et~al.}(2010){Bemporad}, {Soenen}, {Jacobs}, {Landini},
  \& {Poedts}}]{Bemporad2010b}
{Bemporad}, A., {Soenen}, A., {Jacobs}, C., {Landini}, F., \& {Poedts}, S.
  2010, \apj, 718, 251

\bibitem[{{Boe} {et~al.}(2020){Boe}, {Habbal}, {Druckm{\"u}ller}, {Ding},
  {Hod{\'e}rova}, \& {{\v{S}}tarha}}]{Boe2020}
{Boe}, B., {Habbal}, S., {Druckm{\"u}ller}, M., {et~al.} 2020, \apj, 888, 100

\bibitem[{{Boe} {et~al.}(2018){Boe}, {Habbal}, {Druckm{\"u}ller}, {Landi},
  {Kourkchi}, {Ding}, {Starha}, \& {Hutton}}]{Boe2018}
---. 2018, \apj, 859, 155

\bibitem[{Brueckner {et~al.}(1995)Brueckner, Howard, Koomen, Korendyke,
  Michels, Moses, Socker, Dere, Lamy, Llebaria, Bout, Schwenn, Simnett,
  Bedford, \& Eyles}]{Brueckner1995}
Brueckner, G.~E., Howard, R.~a., Koomen, M.~J., {et~al.} 1995, Solar Physics,
  162, 357

\bibitem[{{Cai} {et~al.}(2016){Cai}, {Ning}, \& {Jun}}]{Cai2016}
{Cai}, Qiang-wei, C., {Ning}, W., \& {Jun}, L. 2016, \caa, 40, 352

\bibitem[{{Chen}(1989)}]{Chen1989}
{Chen}, J. 1989, \apj, 338, 453

\bibitem[{{Cheng} {et~al.}(2014){Cheng}, {Ding}, {Zhang}, {Srivastava}, {Guo},
  {Chen}, \& {Sun}}]{Cheng2014a}
{Cheng}, X., {Ding}, M.~D., {Zhang}, J., {et~al.} 2014, \apjl, 789, L35

\bibitem[{{Ciaravella} {et~al.}(2006){Ciaravella}, {Raymond}, \&
  {Kahler}}]{Ciaravella2006}
{Ciaravella}, A., {Raymond}, J.~C., \& {Kahler}, S.~W. 2006, \apj, 652, 774

\bibitem[{{Ciaravella} {et~al.}(2005){Ciaravella}, {Raymond}, {Kahler},
  {Vourlidas}, \& {Li}}]{Ciaravella2005}
{Ciaravella}, A., {Raymond}, J.~C., {Kahler}, S.~W., {Vourlidas}, A., \& {Li},
  J. 2005, \apj, 621, 1121

\bibitem[{{Ciaravella} {et~al.}(2003){Ciaravella}, {Raymond}, {van
  Ballegooijen}, {Strachan}, {Vourlidas}, {Li}, {Chen}, \&
  {Panasyuk}}]{Ciaravella2003}
{Ciaravella}, A., {Raymond}, J.~C., {van Ballegooijen}, A., {et~al.} 2003,
  \apj, 597, 1118

\bibitem[{{Cremades} \& {Bothmer}(2004)}]{Cremades2004}
{Cremades}, H., \& {Bothmer}, V. 2004, \aap, 422, 307

\bibitem[{{Dai} {et~al.}(2014){Dai}, {Wang}, {Huang}, {Du}, \&
  {He}}]{DaiXH2014}
{Dai}, X., {Wang}, H., {Huang}, X., {Du}, Z., \& {He}, H. 2014, \apj, 780, 141

\bibitem[{{Delaboudini{\`e}re} {et~al.}(1995){Delaboudini{\`e}re}, {Artzner},
  {Brunaud}, {Gabriel}, {Hochedez}, {Millier}, {Song}, {Au}, {Dere}, {Howard},
  {Kreplin}, {Michels}, {Moses}, {Defise}, {Jamar}, {Rochus}, {Chauvineau},
  {Marioge}, {Catura}, {Lemen}, {Shing}, {Stern}, {Gurman}, {Neupert},
  {Maucherat}, {Clette}, {Cugnon}, \& {van Dessel}}]{Delab1995}
{Delaboudini{\`e}re}, J.~P., {Artzner}, G.~E., {Brunaud}, J., {et~al.} 1995,
  \solphys, 162, 291

\bibitem[{{Dima} {et~al.}(2018){Dima}, {Kuhn}, {Mickey}, \& {Downs}}]{Dima2018}
{Dima}, G.~I., {Kuhn}, J.~R., {Mickey}, D., \& {Downs}, C. 2018, \apj, 852, 23

\bibitem[{{Feng} {et~al.}(2015){Feng}, {Inhester}, \& {Gan}}]{Feng2015a}
{Feng}, L., {Inhester}, B., \& {Gan}, W. 2015, \apj, 805, 113

\bibitem[{{Feng} {et~al.}(2012){Feng}, {Inhester}, {Wei}, {Gan}, {Zhang}, \&
  {Wang}}]{Feng2012a}
{Feng}, L., {Inhester}, B., {Wei}, Y., {et~al.} 2012, \apj, 751, 18

\bibitem[{{Feng} {et~al.}(2019){Feng}, {Li}, {Chen}, {Li}, {Susino}, {Huang},
  {Lu}, {Ying}, {Li}, {Xue}, {Yang}, {Hong}, {Li}, {Zhao}, {Gan}, \&
  {Zhang}}]{Feng2019}
{Feng}, L., {Li}, H., {Chen}, B., {et~al.} 2019, Research in Astronomy and
  Astrophysics, 19, 162

\bibitem[{{Fineschi} {et~al.}(2020){Fineschi}, {Naletto}, {Romoli}, {Da Deppo},
  {Antonucci}, {Moses}, {Malvezzi}, {Nicolini}, {Spadaro}, {Teriaca},
  {Andretta}, {Capobianco}, {Crescenzio}, {Focardi}, {Frassetto}, {Landini},
  {Massone}, {Melich}, {Nicolosi}, {Pancrazzi}, {Pelizzo}, {Poletto},
  {Sch{\"u}hle}, {Uslenghi}, {Vives}, {Solanki}, {Heinzel}, {Berlicki},
  {Cesare}, {Morea}, {Mottini}, {Sandri}, {Alvarez-Herrero}, \&
  {Castronuovo}}]{Fineschi2020}
{Fineschi}, S., {Naletto}, G., {Romoli}, M., {et~al.} 2020, Experimental
  Astronomy, doi:10.1007/s10686-020-09662-z

\bibitem[{{Frazin} {et~al.}(2002){Frazin}, {Romoli}, {Kohl}, {Gardner}, {Wang},
  {Howard}, \& {Kucera}}]{Frazin2002}
{Frazin}, R.~A., {Romoli}, M., {Kohl}, J.~L., {et~al.} 2002, ISSI Scientific
  Reports Series, 2, 249

\bibitem[{{Gan} {et~al.}(2019){Gan}, {Zhu}, {Deng}, {Li}, {Su}, {Zhang},
  {Chen}, {Zhang}, {Wu}, {Deng}, {Huang}, {Yang}, {Cui}, {Chang}, {Wang}, {Wu},
  {Yin}, {Chen}, {Fang}, {Yan}, {Lin}, {Xiong}, {Chen}, {Bao}, {Cao}, {Bai},
  {Wang}, {Chen}, {Li}, {Zhang}, {Feng}, {Su}, {Li}, {Chen}, {Li}, {Su}, {Wu},
  {Gu}, {Huang}, \& {Tang}}]{Gan2019}
{Gan}, W.-Q., {Zhu}, C., {Deng}, Y.-Y., {et~al.} 2019, Research in Astronomy
  and Astrophysics, 19, 156

\bibitem[{{Gardner} {et~al.}(2002){Gardner}, {Smith}, {Kohl}, {Atkins},
  {Ciaravella}, {Miralles}, {Panasyuk}, {Raymond}, {Strachan}, {Suleiman},
  {Romoli}, \& {Fineschi}}]{Gardner2002}
{Gardner}, L.~D., {Smith}, P.~L., {Kohl}, J.~L., {et~al.} 2002, ISSI Scientific
  Reports Series, 2, 161

\bibitem[{{Gibson} {et~al.}(1999){Gibson}, {Fludra}, {Bagenal}, {Biesecker},
  {del Zanna}, \& {Bromage}}]{Gibson1999}
{Gibson}, S.~E., {Fludra}, A., {Bagenal}, F., {et~al.} 1999, \jgr, 104, 9691

\bibitem[{{Glesener} {et~al.}(2013){Glesener}, {Krucker}, {Bain}, \&
  {Lin}}]{Glesener2013}
{Glesener}, L., {Krucker}, S., {Bain}, H.~M., \& {Lin}, R.~P. 2013, \apjl, 779,
  L29

\bibitem[{{Gopalswamy}(2006)}]{Gopalswamy2006}
{Gopalswamy}, N. 2006, Washington DC American Geophysical Union Geophysical
  Monograph Series, 165, 207

\bibitem[{{Gopalswamy}(2010)}]{Gopalswamy2010}
{Gopalswamy}, N. 2010, in 20th National Solar Physics Meeting, ed.
  I.~{Dorotovic}, Vol.~20, 108--130

\bibitem[{{Gopalswamy} {et~al.}(2005){Gopalswamy}, {Aguilar-Rodriguez},
  {Yashiro}, {Nunes}, {Kaiser}, \& {Howard}}]{Gopalswamy2005}
{Gopalswamy}, N., {Aguilar-Rodriguez}, E., {Yashiro}, S., {et~al.} 2005,
  Journal of Geophysical Research (Space Physics), 110, A12S07

\bibitem[{Gopalswamy {et~al.}(2009)Gopalswamy, Yashiro, Michalek, Stenborg,
  Vourlidas, Freeland, \& Howard}]{Gopalswamy2009}
Gopalswamy, N., Yashiro, S., Michalek, G., {et~al.} 2009, Earth, Moon, and
  Planets, 104, 295

\bibitem[{{Guhathakurta} {et~al.}(1999){Guhathakurta}, {Fludra}, {Gibson},
  {Biesecker}, \& {Fisher}}]{Guhathakurta1999}
{Guhathakurta}, M., {Fludra}, A., {Gibson}, S.~E., {Biesecker}, D., \&
  {Fisher}, R. 1999, \jgr, 104, 9801

\bibitem[{{Handy} {et~al.}(1999){Handy}, {Acton}, {Kankelborg}, {Wolfson},
  {Akin}, {Bruner}, {Caravalho}, {Catura}, {Chevalier}, {Duncan}, {Edwards},
  {Feinstein}, {Freeland }, {Friedlaender}, {Hoffmann}, {Hurlburt},
  {Jurcevich}, {Katz}, {Kelly}, {Lemen}, {Levay}, {Lindgren}, {Mathur},
  {Meyer}, {Morrison}, {Morrison}, {Nightingale}, {Pope}, {Rehse}, {Schrijver},
  {Shine}, {Shing}, {Strong}, {Tarbell}, {Title}, {Torgerson}, {Golub},
  {Bookbinder}, {Caldwell}, {Cheimets}, {Davis}, {Deluca}, {McMullen},
  {Warren}, {Amato}, {Fisher}, {Maldonado}, \& {Parkinson}}]{Handy1999}
{Handy}, B.~N., {Acton}, L.~W., {Kankelborg}, C.~C., {et~al.} 1999, \solphys,
  187, 229

\bibitem[{{Heinzel} {et~al.}(2016){Heinzel}, {Susino}, {Jej{\v{c}}i{\v{c}}},
  {Bemporad}, \& {Anzer}}]{Heinzel2016}
{Heinzel}, P., {Susino}, R., {Jej{\v{c}}i{\v{c}}}, S., {Bemporad}, A., \&
  {Anzer}, U. 2016, \aap, 589, A128

\bibitem[{{Howard} {et~al.}(2008){Howard}, {Moses}, {Vourlidas}, {Newmark},
  {Socker}, {Plunkett}, {Korendyke}, {Cook}, {Hurley}, {Davila}, {Thompson},
  {St Cyr}, {Mentzell}, {Mehalick}, {Lemen}, {Wuelser}, {Duncan}, {Tarbell},
  {Wolfson}, {Moore}, {Harrison}, {Waltham}, {Lang}, {Davis}, {Eyles},
  {Mapson-Menard}, {Simnett}, {Halain}, {Defise}, {Mazy}, {Rochus}, {Mercier},
  {Ravet}, {Delmotte}, {Auchere}, {Delaboudiniere}, {Bothmer}, {Deutsch},
  {Wang}, {Rich}, {Cooper}, {Stephens}, {Maahs}, {Baugh}, {McMullin}, \&
  {Carter}}]{Howard2008}
{Howard}, R.~A., {Moses}, J.~D., {Vourlidas}, A., {et~al.} 2008, \ssr, 136, 67

\bibitem[{{Kimura} \& {Mann}(1998)}]{Kimura1998}
{Kimura}, H., \& {Mann}, I. 1998, Earth, Planets, and Space, 50, 493

\bibitem[{{Kohl} {et~al.}(2006){Kohl}, {Noci}, {Cranmer}, \&
  {Raymond}}]{Kohl2006}
{Kohl}, J.~L., {Noci}, G., {Cranmer}, S.~R., \& {Raymond}, J.~C. 2006, \aapr,
  13, 31

\bibitem[{Kohl {et~al.}(1995)Kohl, Esser, Gardner, Habbal, Daigneau, Dennis,
  Nystrom, Panasyuk, Raymond, Smith, Strachan, Van~Ballegooijen, Noci,
  Fineschi, Romoli, Ciaravella, Modigliani, Huber, Antonucci, Benna, Giordano,
  Tondello, Nicolosi, Naletto, Pernechele, Spadaro, Poletto, Livi, Von
  Der~L{\"u}he, Geiss, Timothy, Gloeckler, Allegra, Basile, Brusa, Wood,
  Siegmund, Fowler, Fisher, \& Jhabvala}]{Kohl1995}
Kohl, J.~L., Esser, R., Gardner, L.~D., {et~al.} 1995, Solar Physics, 162, 313

\bibitem[{{Kohl} {et~al.}(1999){Kohl}, {Esser}, {Cranmer}, {Fineschi},
  {Gardner}, {Panasyuk}, {Strachan}, {Suleiman}, {Frazin}, \&
  {Noci}}]{Kohl1999}
{Kohl}, J.~L., {Esser}, R., {Cranmer}, S.~R., {et~al.} 1999, \apjl, 510, L59

\bibitem[{{Koutchmy} \& {Lamy}(1985)}]{Koutchmy1985}
{Koutchmy}, S., \& {Lamy}, P.~L. 1985, {The F-Corona and the Circum-Solar Dust
  Evidences and Properties (ir)}, ed. R.~H. {Giese} \& P.~{Lamy}, 63

\bibitem[{Lamy {et~al.}(2017)Lamy, Vivès, Curdt, Damé, Davila, Defise,
  Fineschi, Heinzel, Howard, Kuzin, Schmutz, Tsinganos, \& Zhukov}]{Lamy2010}
Lamy, P.~L., Vivès, S., Curdt, W., {et~al.} 2017, in International Conference
  on Space Optics — ICSO 2010, ed. E.~Armandillo, B.~Cugny, \& N.~Karafolas,
  Vol. 10565, International Society for Optics and Photonics (SPIE), 179 -- 185

\bibitem[{{Landi} {et~al.}(2012){Landi}, {Gruesbeck}, {Lepri}, {Zurbuchen}, \&
  {Fisk}}]{Landi2012}
{Landi}, E., {Gruesbeck}, J.~R., {Lepri}, S.~T., {Zurbuchen}, T.~H., \& {Fisk},
  L.~A. 2012, \apj, 761, 48

\bibitem[{Lanzerotti(2013)}]{Lanzerotti2001}
Lanzerotti, L.~J. 2013, Space Weather Effects on Technologies (American
  Geophysical Union (AGU)), 11--22

\bibitem[{{Lemaire} {et~al.}(2002){Lemaire}, {Emerich}, {Vial}, {Curdt},
  {Sch{\"u}hle}, \& {Wilhelm}}]{Lemaire2002}
{Lemaire}, P., {Emerich}, C., {Vial}, J.-C., {et~al.} 2002, in ESA Special
  Publication, Vol. 508, From Solar Min to Max: Half a Solar Cycle with SOHO,
  ed. A.~{Wilson}, 219--222

\bibitem[{{Lemen} {et~al.}(2012){Lemen}, {Title}, {Akin}, {Boerner}, {Chou},
  {Drake}, {Duncan}, {Edwards}, {Friedlaender}, {Heyman}, {Hurlburt}, {Katz},
  {Kushner}, {Levay}, {Lindgren}, {Mathur}, {McFeaters}, {Mitchell}, {Rehse},
  {Schrijver}, {Springer}, {Stern}, {Tarbell}, {Wuelser}, {Wolfson}, {Yanari},
  {Bookbinder}, {Cheimets}, {Caldwell}, {Deluca}, {Gates}, {Golub}, {Park},
  {Podgorski}, {Bush}, {Scherrer}, {Gummin}, {Smith}, {Auker}, {Jerram},
  {Pool}, {Soufli}, {Windt}, {Beardsley}, {Clapp}, {Lang}, \&
  {Waltham}}]{Lemen2012}
{Lemen}, J.~R., {Title}, A.~M., {Akin}, D.~J., {et~al.} 2012, \solphys, 275, 17

\bibitem[{{Li} {et~al.}(2019){Li}, {Chen}, {Feng}, {Li}, {Huang}, {Li}, {Lu},
  {Xue}, {Ying}, {Zhao}, {Yang}, {Gan}, {Fang}, {Song}, {Wang}, {Guo}, {He},
  {Zhu}, {Zhu}, {Deng}, {Bao}, {Cao}, \& {Yang}}]{Li2019}
{Li}, H., {Chen}, B., {Feng}, L., {et~al.} 2019, Research in Astronomy and
  Astrophysics, 19, 158

\bibitem[{{Liu} {et~al.}(2015){Liu}, {Wang}, {Shen}, {Liu}, {Pan}, \&
  {Wang}}]{Liujiajia2015}
{Liu}, J., {Wang}, Y., {Shen}, C., {et~al.} 2015, \apj, 813, 115

\bibitem[{{Lu} {et~al.}(2017){Lu}, {Inhester}, {Feng}, {Liu}, \&
  {Zhao}}]{Lu2017}
{Lu}, L., {Inhester}, B., {Feng}, L., {Liu}, S., \& {Zhao}, X. 2017, \apj, 835,
  188

\bibitem[{{Miralles} {et~al.}(2001){Miralles}, {Cranmer}, {Panasyuk}, {Romoli},
  \& {Kohl}}]{Miralles2001}
{Miralles}, M.~P., {Cranmer}, S.~R., {Panasyuk}, A.~V., {Romoli}, M., \&
  {Kohl}, J.~L. 2001, \apjl, 549, L257

\bibitem[{{Moran} \& {Davila}(2004)}]{Moran2004}
{Moran}, T.~G., \& {Davila}, J.~M. 2004, Science, 305, 66

\bibitem[{{Morgan} {et~al.}(2012){Morgan}, {Byrne}, \& {Habbal}}]{Morgan2012}
{Morgan}, H., {Byrne}, J.~P., \& {Habbal}, S.~R. 2012, \apj, 752, 144

\bibitem[{{Morgan} \& {Habbal}(2007)}]{Morgan2007}
{Morgan}, H., \& {Habbal}, S.~R. 2007, \aap, 471, L47

\bibitem[{{Noci} {et~al.}(1987){Noci}, {Kohl}, \& {Withbroe}}]{Noci1987}
{Noci}, G., {Kohl}, J.~L., \& {Withbroe}, G.~L. 1987, \apj, 315, 706

\bibitem[{{Ofman} {et~al.}(1997){Ofman}, {Romoli}, {Poletto}, {Noci}, \&
  {Kohl}}]{Ofman1997}
{Ofman}, L., {Romoli}, M., {Poletto}, G., {Noci}, G., \& {Kohl}, J.~L. 1997,
  \apjl, 491, L111

\bibitem[{{Ofman} {et~al.}(2000){Ofman}, {Romoli}, {Poletto}, {Noci}, \&
  {Kohl}}]{Ofman2000}
---. 2000, \apj, 529, 592

\bibitem[{{Pagano} {et~al.}(2020){Pagano}, {Bemporad}, \&
  {Mackay}}]{Pagano2020}
{Pagano}, P., {Bemporad}, A., \& {Mackay}, D.~H. 2020, \aap, 637, A49

\bibitem[{{Panesar} {et~al.}(2016){Panesar}, {Sterling}, \&
  {Moore}}]{Panesar2016}
{Panesar}, N.~K., {Sterling}, A.~C., \& {Moore}, R.~L. 2016, \apjl, 822, L23

\bibitem[{Prasad {et~al.}(2017)Prasad, Banerjee, Singh, Nagabhushana, Kumar,
  Kamath, Kathiravan, Venkata, Rajkumar, Natarajan, Juneja, Somu, Pant, Shaji,
  Sankarsubramanian, Patra, Venkateswaran, Adoni, Narendra, \&
  Haridas}]{Prasad2017}
Prasad, B.~R., Banerjee, D., Singh, J., {et~al.} 2017, CURRENT SCIENCE, 113,
  613

\bibitem[{{Qu{\'e}merais} \& {Lamy}(2002)}]{Quemerais2002}
{Qu{\'e}merais}, E., \& {Lamy}, P. 2002, \aap, 393, 295

\bibitem[{{Ragot} \& {Kahler}(2003)}]{Ragot2003}
{Ragot}, B.~R., \& {Kahler}, S.~W. 2003, \apj, 594, 1049

\bibitem[{{Raymond} {et~al.}(2003){Raymond}, {Ciaravella}, {Dobrzycka},
  {Strachan}, {Ko}, {Uzzo}, \& {Raouafi}}]{Raymond2003}
{Raymond}, J.~C., {Ciaravella}, A., {Dobrzycka}, D., {et~al.} 2003, \apj, 597,
  1106

\bibitem[{{Raymond} {et~al.}(1997){Raymond}, {Kohl}, {Noci}, {Antonucci},
  {Tondello}, {Huber}, {Gardner}, {Nicolosi}, {Fineschi}, {Romoli}, {Spadaro},
  {Siegmund}, {Benna}, {Ciaravella}, {Cranmer}, {Giordano}, {Karovska},
  {Martin}, {Michels}, {Modigliani}, {Naletto}, {Panasyuk}, {Pernechele},
  {Poletto}, {Smith}, {Suleiman}, \& {Strachan}}]{Raymond1997}
{Raymond}, J.~C., {Kohl}, J.~L., {Noci}, G., {et~al.} 1997, \solphys, 175, 645

\bibitem[{{Reames}(2013)}]{Reames2013}
{Reames}, D.~V. 2013, \ssr, 175, 53

\bibitem[{{Reeves} \& {Golub}(2011)}]{Reeves2011}
{Reeves}, K.~K., \& {Golub}, L. 2011, \apjl, 727, L52

\bibitem[{{Romoli} {et~al.}(2002){Romoli}, {Frazin}, {Kohl}, {Gardner},
  {Cranmer}, {Reardon}, \& {Fineschi}}]{Romoli2002}
{Romoli}, M., {Frazin}, R.~A., {Kohl}, J.~L., {et~al.} 2002, ISSI Scientific
  Reports Series, 2, 181

\bibitem[{{Schrijver} \& {De Rosa}(2003)}]{Schrijver2003}
{Schrijver}, C.~J., \& {De Rosa}, M.~L. 2003, \solphys, 212, 165

\bibitem[{Schwenn(2006)}]{Schwenn2006}
Schwenn, R. 2006, Living Reviews in Solar Physics, 3, 2

\bibitem[{Susino \& Bemporad(2016)}]{Susino2016}
Susino, R., \& Bemporad, A. 2016, The Astrophysical Journal, 830, 58

\bibitem[{{Susino} {et~al.}(2014){Susino}, {Bemporad}, \& {Dolei}}]{Susino2014}
{Susino}, R., {Bemporad}, A., \& {Dolei}, S. 2014, \apj, 790, 25

\bibitem[{{Susino} {et~al.}(2018){Susino}, {Bemporad}, {Jej{\v{c}}i{\v{c}}}, \&
  {Heinzel}}]{Susino2018}
{Susino}, R., {Bemporad}, A., {Jej{\v{c}}i{\v{c}}}, S., \& {Heinzel}, P. 2018,
  \aap, 617, A21

\bibitem[{{Tritschler} {et~al.}(2015){Tritschler}, {Rimmele}, {Berukoff},
  {Casini}, {Craig}, {Elmore}, {Hubbard}, {Kuhn}, {Lin}, {McMullin}, {Reardon},
  {Schmidt}, {Warner}, \& {Woger}}]{Tritschler2015}
{Tritschler}, A., {Rimmele}, T.~R., {Berukoff}, S., {et~al.} 2015, in Cambridge
  Workshop on Cool Stars, Stellar Systems, and the Sun, Vol.~18, 18th Cambridge
  Workshop on Cool Stars, Stellar Systems, and the Sun, 933--944

\bibitem[{{van de Hulst}(1950)}]{VandeHulst1950}
{van de Hulst}, H.~C. 1950, \bain, 11, 135

\bibitem[{{Vourlidas} {et~al.}(2003){Vourlidas}, {Wu}, {Wang}, {Subramanian},
  \& {Howard}}]{Vourlidas2003}
{Vourlidas}, A., {Wu}, S.~T., {Wang}, A.~H., {Subramanian}, P., \& {Howard},
  R.~A. 2003, \apj, 598, 1392

\bibitem[{{Wang} {et~al.}(2009){Wang}, {Zhang}, \& {Shen}}]{Wang2009}
{Wang}, Y., {Zhang}, J., \& {Shen}, C. 2009, Journal of Geophysical Research
  (Space Physics), 114, A10104

\bibitem[{Webb \& Howard(2012)}]{Webb2012}
Webb, D.~F., \& Howard, T.~A. 2012, Living Reviews in Solar Physics, 9, 3

\bibitem[{{Withbroe} {et~al.}(1982){Withbroe}, {Kohl}, {Weiser}, \&
  {Munro}}]{Withbroe1982}
{Withbroe}, G.~L., {Kohl}, J.~L., {Weiser}, H., \& {Munro}, R.~H. 1982, \ssr,
  33, 17

\bibitem[{{Ying} {et~al.}(2019){Ying}, {Bemporad}, {Giordano}, {Pagano},
  {Feng}, {Lu}, {Li}, \& {Gan}}]{Ying2019}
{Ying}, B., {Bemporad}, A., {Giordano}, S., {et~al.} 2019, \apj, 880, 41

\bibitem[{{Ying} {et~al.}(2018){Ying}, {Feng}, {Lu}, {Zhang}, {Magdalenic},
  {Su}, {Su}, \& {Gan}}]{Ying2018}
{Ying}, B., {Feng}, L., {Lu}, L., {et~al.} 2018, \apj, 856, 24

\bibitem[{{Zhang} {et~al.}(2012){Zhang}, {Cheng}, \& {Ding}}]{Zhangjie2012}
{Zhang}, J., {Cheng}, X., \& {Ding}, M.-D. 2012, Nature Communications, 3, 747

\end{thebibliography}
\end{document}